\documentclass[12pt,preprint]{aastex}
\usepackage{epsfig}
\usepackage{natbib}
\usepackage{graphicx}
\usepackage{slashbox}
\usepackage{multirow}
\usepackage{lscape}
\usepackage{mathrsfs,amssymb}
\usepackage{amsmath}
\usepackage{subfigure}
\usepackage{amssymb}
\usepackage{multirow}
\usepackage{tabularx}
\usepackage{rotating}
\usepackage{amsmath}
\usepackage{ulem}
\usepackage{lineno}

\newcommand       \cm           {\,{\rm cm}}

\newcommand       \K            {\,{\rm K}}

\newcommand       \yr       {\,{\rm yr}}

\newcommand       \NH           {N_{\rm H}}
\newcommand       \simlt        {\lesssim}
\newcommand       \simgt        {\gtrsim}

\newcommand       \mum          {\,{\rm \mu m}}
\newcommand       \ppm          {\,{\rm ppm}}

\newcommand       \simali       {\sim\,}

\newcommand \dprim {\left[\rm D/H\right]_{\rm prim}}
\newcommand \Dism {\left[\rm D/H\right]_{\rm ISM}}
\newcommand \dism {\left[\rm D/H\right]_{\rm ISM}}

\newcommand \dgas {\left[\rm D/H\right]_{\rm gas}}

\newcommand \Dpah {\left[\rm D/H\right]_{\scriptscriptstyle\rm PAH}}

\newcommand       \Aratio        {A_{4.4}/A_{3.3}}
\newcommand       \Acd           {A_{4.4}}
\newcommand       \ACD           {A_{4.4}}
\newcommand       \Aaro          {A_{3.3}}
\newcommand       \ACH          {A_{3.3}}
\newcommand       \NC         {N_{\rm C}}
\newcommand       \ND         {{N_{\rm D}}}
\newcommand       \Iratioobs     {\left(I_{3.3}/I_{4.4}\right)_{\rm obs}}

\newcommand       \km        {\,{\rm km}}

\newcommand       \mol       {\,{\rm mol}}

\countdef\decade=200
\advance\decade by \year
\countdef\hours=201
\advance\hours by \time
\divide\hours by 60
\countdef\mins=202
\advance\mins by \hours
\multiply\mins by 60
\multiply\hours by 100
\countdef\miltime=203
\advance\miltime by \hours
\advance\miltime by \time
\advance\miltime by -\mins
\def\today{\number\decade.\number\month.\number\day.\number\miltime}

\shorttitle{IR spectra of Multi-Deuterated PAHs}
\title{
Deuterated Polycyclic Aromatic Hydrocarbons
in the Interstellar Medium:
The C--D Band Strengths of Multi-Deuterated Species
\\{\small DRAFT: \today ~~}
}
\author{X.J.~Yang\altaffilmark{1,2},
  Aigen Li\altaffilmark{2},
  C.Y.~He\altaffilmark{3},
            and R.~Glaser\altaffilmark{4}
            }
            \altaffiltext{1}{Hunan Key Laboratory for Stellar
              and Interstellar Physics
              and School of Physics and Optoelectronics,
              Xiangtan University, Hunan 411105, China;
                      {\sf xjyang@xtu.edu.cn}}
\altaffiltext{2} {Department of Physics and Astronomy,
                  University of Missouri,
                  Columbia, MO 65211, USA;
                  {\sf lia@missouri.edu}}
\altaffiltext{3}{Hunan Key Laboratory for
                  Micro-Nano Energy Materials and Devices
              and School of Physics and Optoelectronics,
              Xiangtan University, Hunan 411105, China}
\altaffiltext{4} {Department of Chemistry,
                  Missouri University of Science and Technology,
                  Rolla, MO 65409, USA}

\begin{document}

\begin{abstract}
  Observationally, the interstellar gas-phase abundance
  of deuterium (D) is considerably depleted and
  the missing D atoms are often postulated to
  have been locked up into carbonaceous solids
  and polycyclic aromatic hydrocarbon (PAH) molecules.
  An accurate knowledge of the fractional amount of D
  (relative to H) tied up in carbon dust and PAHs
  has important cosmological implications
  since D originated exclusively from the Big Bang
  and the present-day D abundance, after accounting
  for the astration it has experienced
  during the Galactic evolution, provides essential clues
  to the primordial nucleosynthesis
  and the cosmological parameters.
  To quantitatively explore the extent to which PAHs
  could possibly accommodate the observed D depletion,
  we have previously quantum-chemically computed
  the infrared vibrational spectra
  of {\it mono-deuterated} PAHs
  and derived  the mean intrinsic band
  strengths of the 3.3$\mum$ C--H stretch ($\ACH$)
  and the 4.4$\mum$ C--D stretch ($\ACD$).
  Here we extend our previous work to
  {\it multi-deuterated} PAH species of
  different deuterations, sizes and structures.
  We find that both the intrinsic band strengths
  $\ACH$ and $\ACD$
  and their ratios
  $\Aratio$ not only show little variations among PAHs
  of different deuterations, sizes and structures,
  they are also closely similar to that of mono-deuterated PAHs.
  Therefore, a PAH deuteration level
  (i.e., the fraction of peripheral atoms
attached to C atoms in the form of D)
of $\simali$2.4\% previously estimated
from the observed 4.4$\mum$ to 3.3$\mum$
band ratio based on the $\Aratio$ ratio of
{\it mono-deuterated} PAHs is robust.
\end{abstract}


\section{Introduction\label{sec:intro}}

The interstellar abundance of deuterium (D) provides
important insights into the Big Bang cosmology.
Exclusively created in the Big Bang, the primordial
abundance of D depends sensitively on
the cosmological parameters such as
the baryon closure parameter $\Omega_b$
and the Hubble constant $h$
(see Boesgaard \& Steigman 1985).
The standard Big Bang Nucleosynthesis (BBN) model
predicts a primordial D/H abundance of
 $\dprim$\,$\approx$\,26$\ppm$
from the $\Omega_b$ and $h$ parameters
derived from observations of
the cosmic microwave background
(e.g., see Spergel et al.\ 2003,
Coc et al.\ 2004, S{\'a}nchez et al.\ 2006).
This is closely consistent with
the observationally determined abundance
of $\dprim$\,$\approx$\,25--28$\ppm$,
based on high precision measurements
of the column densities of D\,{\sc i}
and H\,{\sc i} of compositionally ``pristine''
metal-poor quasar absorption line systems
and damped Ly$\alpha$ absorption systems
(e.g., see Cooke et al.\ 2018, Zavarygin et al.\ 2018).
%
%

Once incorporated into stars,
deuterium could be easily destroyed by
nuclear fushion in stellar interiors,
a process known as ``astration''
which converts D to $^3$He, $^4$He,
and heavier elements.
Conventional Galactic chemical evolution models
predict a monotonic decrease in D/H with time
(Mazzitelli \& Moertti 1980),
with a present-day D/H abundance of
$\dism$\,$\simgt$\,$20\pm1\ppm$
in the Galactic interstellar medium
(ISM; see Prodanovi{\'c} et al.\ 2010).
However, high resolution ultraviolet (UV)
spectroscopic observations of the local ISM
have revealed that the interstellar gas-phase
$\dgas$ abundance exhibits
substantial regional variations
(by a factor of $\simali$4)
within a few hundred parsecs of the Sun,
with some lines of sight exhibiting
a much lower gas-phase D/H abundance
than the expected interstellar D/H
abundance of
$\dism$\,$\simgt$\,$20\pm1\ppm$
(see Draine 2006 and referenes therein).\footnote{%
         Using the UV absorption spectra
          obtained with the Copernicus satellite,
          Allen et al.\ (1992) derived
          $\dgas$\,=\,$5.0\pm1.6\ppm$
          toward $\theta$~Car
          at a distance of $d=135\pm9\,$pc.
          Jenkins et al.\ (1999) measured
          $\dgas$\,=\,$7.4_{-0.9}^{+1.2}\ppm$
          toward $\delta$~Ori ($d=281\pm65$\,pc),
          based on the Ly$\delta$ and Ly$\varepsilon$
          absorption features of D
          obtained with the Interstellar Medium
          Absorption Profile Spectrograph
          (IMAPS) on the ORFEUS-SPAS II mission.
          }
          %



The puzzling underabundance of
the gas-phase $\dgas$ abundance
(in comparison with the present-epoch
interstellar D/H abundance
of $\dism$\,$\simgt$\,$20\pm1\ppm$
expected from the BBN model and
the Galactic chemical evolution model)
and the apparent regional
variations of $\dgas$
could be related to carbonaceous solids
and polycyclic aromatic hydrocarbon (PAH) molecules.
%
More specifically, Draine (2004) proposed that
D might be depleted from the gas phase
and sequestered in PAHs.
The variations in the gas-phase $\dgas$ abundance
could be attributed to variations from one sightline
to another in the fraction of the D atoms
sequestered in PAHs.
Many processes have been proposed to
drive deuterium enrichment in PAHs
(see Sandford et al.\ 2001).
Due to the lower zero-point energy
of C--D bonds
(in comparison with C--H bonds),
PAHs of intermediate size are expected
to become deuterium enriched in the ISM
through the selective loss of hydrogen
during photodissociation events
(Allamandola et al.\ 1989).
Draine (2006) also showed that collisions
of D with PAH cations and D$^+$ with PAHs
could incorporate D into PAHs
and efficiently deplete D from the gas
on time scales of $\simali$2\,Myr
in cool diffuse clouds.
%
Bernstein  et  al.\ (1999) and
Sandford et al.\ (2000) have demonstrated
experimentally that UV photolysis of PAHs
in D-enriched ices could result in rapid
D enrichment and even produce
multi-deuterated PAHs.
The latter process is particularly relevant
for dense regions where ices are photochemically
processed by ambient UV photons.
Deuterium  enrichments of extraterrestrial
organic compounds have been experimentally
measured, including aromatics in meteorites
and interplanetary dust particles
(Allamandola et al.\ 1987,
Kerridge et al.\ 1987,
Clemett et al.\ 1993).
Indeed, PAHs represent a major carrier
of D excesses in primitive meteorites
(Kerridge et al.\ 1987).

PAHs are abundant and widespread
throughout the Universe, as revealed by their
distinctive set of infrared (IR) emission bands
at 3.3, 6.2, 7.7, 8.6, 11.3 and 12.7$\mum$,
which are characteristic of their C--H and
C--C vibrational modes and often also known
as the ``unidentified infrared emission''
(IUE) bands (see Li 2020).\footnote{%
    In astronomy, ``PAH'' is a generic term.
    To fit the observed astronomical UIE spectra,
    the PAH model requires a mixture of
    {\it specific} PAH molecules
    (e.g., see Cami 2011, Rosenberg et al.\ 2011)
    or empirical ``astro-PAHs''
    (Li \& Draine 2001, Draine \& Li 2007, Draine et al.\ 2021)
    of various sizes and chargings states
    (i.e., neutrals, cations and anions).
    No specific PAH molecules have been
    identified in space until  
    very recently McGuire et al.\ (2021)
    reported the first detection of two isomers
    of cyanonapthalene (C$_{11}$H$_7$N),
    a bicyclic ring molecule,
    in the TMC-1 molecular cloud.
    }
Upon deuteration (i.e., one or more peripheral
H atoms are replaced by D), additional bands
at $\simali$4.4$\mum$ (C--D stretching),
$\simali$11.7$\mum$ (C--D in-plane bending),
and $\simali$15.4$\mum$
(C--D out-of-plane [oop] bending) are expected
to show up in their emission spectra
(see Hudgins et al.\ 1994, 2004,
Bauschlicher et al.\ 1997,
Draine 2006, Yang et al.\ 2021b).
While the 11.7$\mum$ C--D band
might be confused with
the C--H oop bending modes
at $\simali$11.3$\mum$
and the 15.4$\mum$ C--D band
falls in a region where various weak
C--C skeleton modes are present,
the 4.4$\mum$ C--D band appears
to be the best probe of the deuteration
of interstellar PAHs.
Indeed, a weak band at 4.4$\mum$ has been
detected by the Infrared Space Observatory
and AKARI in photodissociated regions,
reflection nebulae, and H\,{\sc ii} regions
in both the Milky Way and Large and Small
Magellanic Clouds (Peeters et al.\ 2004,
Onaka et al.\ 2014, Doney et al.\ 2016).
Meanwhile, a weak band at 4.65$\mum$
was also detected in these regions,
which is generally attributed to emission
from aliphatic C--D stretch.

To quantitatively explore the hypothesis of
PAHs as a possible reservoir of interstellar D,
we have previously performed quantum-chemical
computations on the IR vibrational spectra of
{\it mono-deuterated} PAHs of various sizes
and determined the intrinsic band strength
of the 4.4$\mum$ C--D stretch.
By comparing the computed 4.4$\mum$ C--D
band strength with the observed intensities,
we have derived the degree of deuteration
(i.e., the fraction of peripheral atoms
attached to C atoms in the form of D)
to be $\simali$2.4\%, implying that interstellar
PAHs could be D-enriched by a factor of
$\Dpah/\Dism\approx1200$
(Yang et al.\ 2020b).
However, it is not impossible that some PAHs
may be {\it multi-deuterated} in the ISM,
i.e., two or more peripheral H atoms are
replaced by D atoms.
The detection of various multiply-deuterated
gas-phase species in dense clouds
such as D$_2$CO (Loinard et al.\ 2000),
NHD$_2$ and ND$_3$
(Roueff et al.\ 2000, van der Tak et al.\ 2002),
CHD$_2$OH (Parise et al.\ 2002),
and D$_2$S (Vastel et al.\ 2003)
indicates that multiple deuteration
may also occur to interstellar PAHs.
%
To explore the photo-chemistry of small PAHs
in ices under interstellar conditions,
Bernstein et al.\ (1999) and Sandford et al.\ (2000)
applied UV irradiation to PAH-containing D-rich ices
and found that UV photolysis of PAHs
in low-temperature ices could easily cause
one or more of the peripheral H atoms on PAHs
to be replaced by D atoms.
These deuterated and multi-deuterated PAHs
created in dense clouds
could be present in the diffuse ISM
through rapid exchange of material
between diffuse and dense clouds.\footnote{%
  It is well recognized that in the diffuse ISM,
  dust grains (e.g., silicate and graphite)
  are destroyed at a rate faster than their
  stellar production (McKee 1989).
  This led Draine (1990) to conclude that
  the bulk of the solid material
  in grains actually condensed
  in the ISM rather than in stellar outflows.
  Draine (1990) argued that there must be rapid
  exchange of matter between the diffuse ISM
  and molecular clouds on a timescale of
  $\simali$$2\times10^7\yr$ or less
  since the bulk of grain growth can proceed
  rapidly only in dense regions.
  }
To complement our previous work
on mono-deuterated PAHs, in this work
we focus on multi-deuterated species,
aiming at deriving the intrinsic band strength of
the 4.4$\mum$ C--D stretch of multi-deuterated
PAHs and comparing it with that of mono-deuterated
species. To this end, we perform quantum-chemical
computations on the IR vibrational spectra of
a large sample of multi-deuterated molecules.
This paper is organized as follows.
In \S\ref{sec:method} we briefly describe
the computational methods and
the selected target molecules.
The computed IR vibrational spectra
and the derived intrinsic C--D band strengths
are reported in \S\ref{sec:results}.
In \S\ref{sec:astro} we present
the recommended C--D band strengths
and discuss the astrophysical implications.
Finally, we summarize our major results
in \S\ref{sec:summary}.

\section{Computational Methods
         and Target Molecules
         \label{sec:method}
         }
\begin{figure*}
\centering{
\includegraphics[scale=0.65,clip]{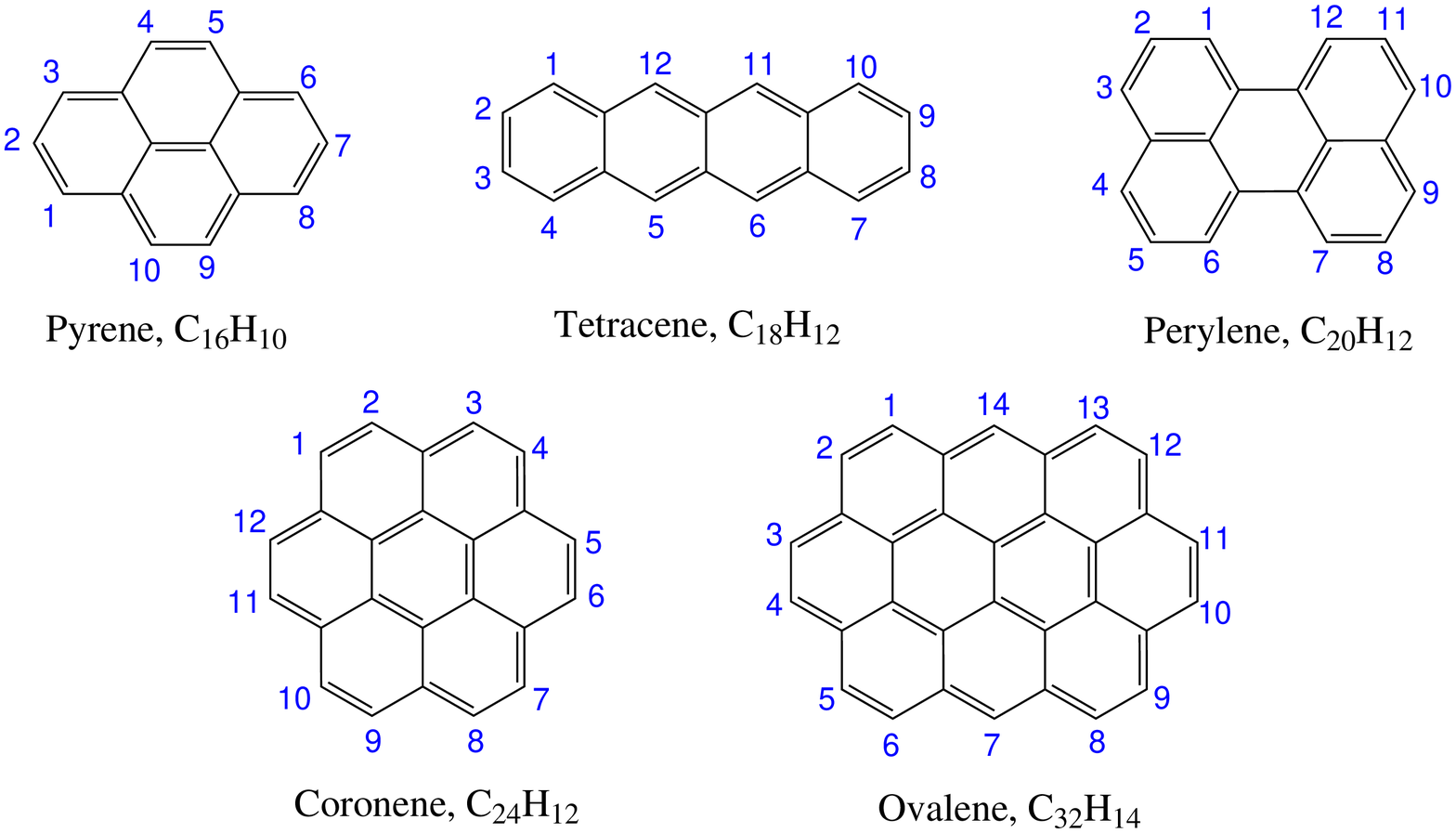}
}
\caption{\footnotesize
         \label{fig:PAH-MultiD_structures}
         Structures of parent PAHs considered here
         for deuteration.
         For a multi-deuterated species,
         we refer it by the abbreviation of
         the first four letters of the name
         of its parental molecule, followed by
         the number of D atoms attached
         (e.g., Pyre$\_$2D refers to
         doubly-deuterated pyrene).
         }
\end{figure*}

We use the Gaussian16 software
(Frisch et al.\ 2016) to calculate
the vibrational spectra for a large
number of multiply-deuterated
species of five parental molecules
(see Figure~\ref{fig:PAH-MultiD_structures}):
pyrene (C$_{16}$H$_{10}$),
tetracene (C$_{18}$H$_{12}$),
perylene (C$_{20}$H$_{12}$),
coronene (C$_{24}$H$_{12}$)
and ovalene (C$_{32}$H$_{14}$).
Our target molecules are mostly
compact and pericondensed
(i.e., pyrene, perylene, coronene, and ovalene).
For comparison, we also consider
one linear, catacondensed molecule
(i.e., tetracene).
Larger PAHs of more than 40 carbon atoms
(i.e., $\NC\simgt40$) are less likely deuterated
(see Hudgins et al.\ 2004)
since they have a large number of internal degrees
of freedom to accommodate the absorbed photon
energy and hence perypheral C--H bond rupture
does not occur.

For each parent PAH molecule,
we consider all deuteration possibilites,
from mono-deuteration\footnote{%
  For completeness, the band strengths
  computed for mono-deuterated species
  reported in Yang et al.\ (2020b) are also
  included here when we derive the mean
  band strengths for all isomers.
  Note that mono-deuterated isomers
  are only a minor fraction of our target sample.
  The inclusion or exclusion of mono-deuterated
  isomers essentially does not affect our results.
  }
all the way up to complete deuteration
or perdeuteration.\footnote{%
  Perdeuterated PAHs refer to PAHs
  in which all of the peripheral H atoms
  have been replaced by D atoms.
  }
We assume that all the D atoms are attached
to the aromatic C atoms,
although both astronomical observations
and UV irradiation of PAHs in interstellar
ice analogs leads have suggested that
D could also be attached to aliphatic C atoms
(Bernstein et al.\ 1999,
Sandford et al.\ 2000, 2001,
Peeters et al.\ 2004,
Onaka et al.\ 2014, Doney et al.\ 2016,
Buragohain et al.\ 2015, 2016).

\begin{figure*}
\centering{
\includegraphics[scale=0.75,clip]{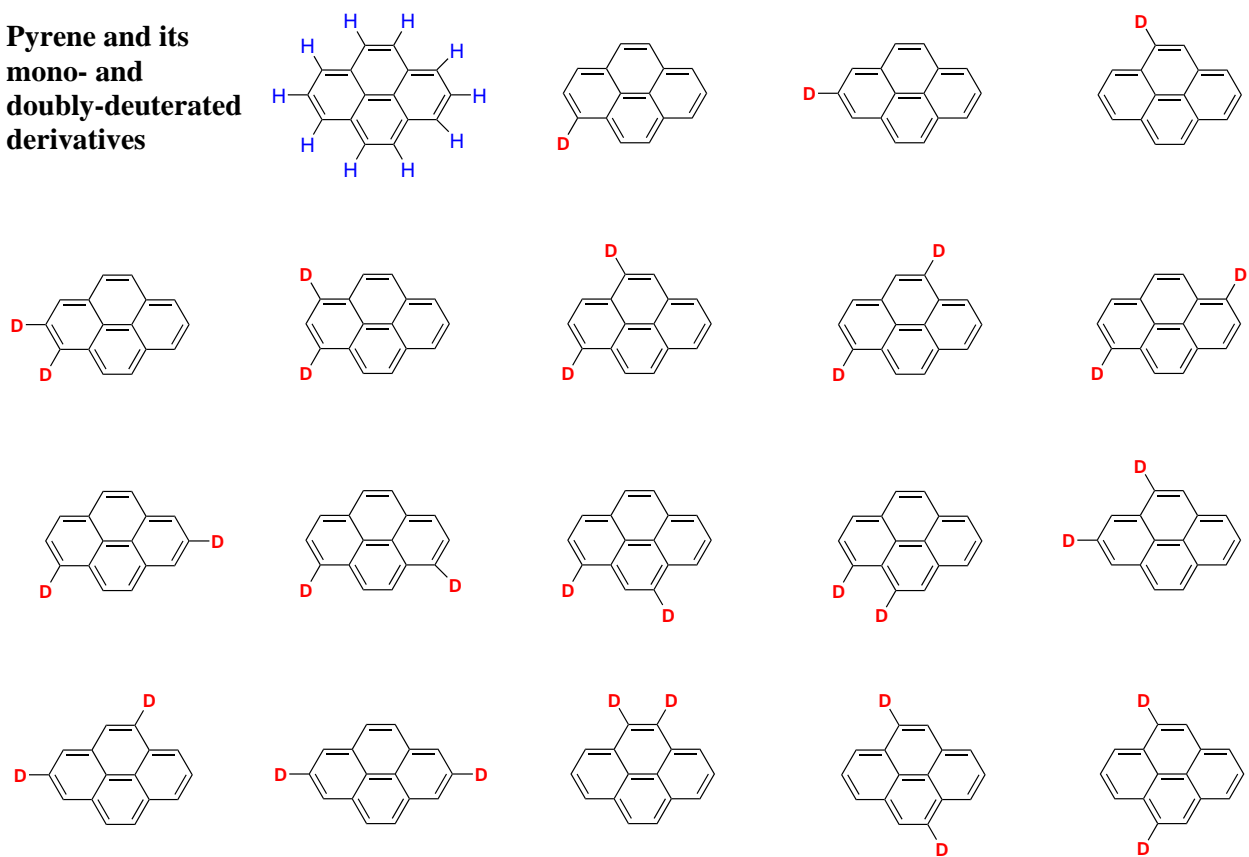}
}
\caption{\footnotesize
         \label{fig:Pyrene_2D_structures}
         Structures of pyrene (C$_{16}$H$_{10}$)
         and its three mono-deuterated isomers
         (Pyre$\_$1D)
         and 15 di-deuterated isomers (Pyre$\_$2D).
         }
\end{figure*}

Similar to Yang et al.\ (2020b),
we refer mono-deuterated species
by the abbreviation of the first four letters of
the names of their parental PAH molecules
followed by the position where the D atom
is attached (e.g., Pyre$\_$D2 refers to
mono-deuterated pyrene in which the (only)
D atom is attached at position 2).
For multiply-deuterated species,
we refer them also by the abbreviation of
the first four letters of the names of their
parental molecules, but followed by
the number of D atoms attached
(e.g., Pyre$\_$2D refers to
di-deuterated pyrene).
For each multi-deuterated molecule
(of the same degree of deuteration),
depending on the positioning
of the D atoms, there are many isomers
(e.g., Pyre$\_$2D has 15 isomers;
see Figure~\ref{fig:Pyrene_2D_structures}).\footnote{%
  Consider a PAH molecule consisting of
  $x$ carbon atoms and $y$ hydrogen atoms
  (i.e., C$_{x}$H$_{y}$).
  With $z$ peripheral H atoms
  replaced by D atoms,
  it is deuterated into
  C$_{x}$H$_{y-z}$D$_{z}$.
  For each configuration of C$_{x}$H$_{y}$,
  there are in total $2^y$ isomers
  for C$_{x}$H$_{y-z}$D$_{z}$,
  according to the principle
  of permutation and combination.
  However, a large fraction of these isomers
  are equivalent due to symmetry.
  In this work, to avoid duplicate isomers,
  we will confine ourselves
  to those ``inequivalent'' isomers
  given by the RG2 code (Shi et al.\ 2018).
  }
In naming, we do not further distinguish
these isomers. We do distinguish them
when we compute their vibrational spectra
(see \S\ref{sec:results}).
%

Following Yang et al.\ (2020b),
the hybrid density functional theoretical
method (B3LYP) at the {\rm 6-311+G$^{\ast\ast}$}
level is employed for the calculation
of IR vibrational spectra
under the harmonic approximation,
which provides sufficient computational accuracies
with operable computer time
(see Yang et al.\ 2017 and reference therein).
We present the harmonic vibrational frequencies
and intensities, with the standard scaling
applied to the frequencies by employing
a scale factor of $\simali$0.9688 (Borowski 2012).

\section{Results\label{sec:results}}
\subsection{Theoretical Background and
                    Pre-computational Expectations
                    \label{subsec:BO}}
The B3LYP DFT computational technique adopted
in this work is based on the Born-Oppenheimer (BO)
approximation, a simplified description of the quantum
states of molecules which separates the motion of
the nuclei and the motion of the electrons.
The physical basis for the BO approximation is the fact
that the mass of an atomic nucleus in a molecule is much
larger (by over 1000 times) than the mass of an electron.
Because of this mass difference, the nuclei move much
more slowly than the electrons, and the electrons adjust 
instantaneously to any nuclear motion.
%

In the harmonic approximation,      
the vibrational frequency
of a chemical bond
can be estimated from
$\nu=\sqrt{k/\mu}/2\pi$,
where $\mu$ is the reduced mass,
and $k$, the force constant, is the second
derivative of the potential energy $V$,
$k=\partial^2V/\partial Q^2$,
where $Q$ is the vibrational normal coordinate.
The vibrational band intensity $A$ is related to
electron charge displacements
(i.e., changes in the dipole moment vector
${\bf \overrightarrow{\mu}}$
with vibrational coordinate)
occurring in a chemical bond
over the lifetime of a vibration,
$A\propto\left|\partial {\bf
\overrightarrow{\mu}}/\partial Q\right|^2$
(see Wexler 1967).
%
%
Under the BO approximation,
the presence of D atoms only affects
the reduced mass $\mu$ and barely
affects the force constants $k$ and
the dipole moment derivatives
$\partial {\bf\overrightarrow{\mu}}/\partial Q$.
Therefore, one would expect that
the frequency and intensity of
the C--D stretch would not be affected
much by the neighbouring atoms,
i.e., essentially insensitive to the PAH
size and structure as well as the presence
of additional D atoms.
Similarly, the frequency and intensity of
the C--H stretch would also not be affected
much by the presence of D atoms.
The C--H and C--D stretches are expected
to fall at very different frequencies due to
the differences in mass and are not expected
to mix, with the latter occurring at a wavelength
longer by a factor of $\simali$$\sqrt{13/7}$.
%
%

\subsection{Multi-Deuterated Pyrene
                    \label{sec:pyrene}}
\begin{figure*}
\centering{
\includegraphics[scale=0.5,clip]{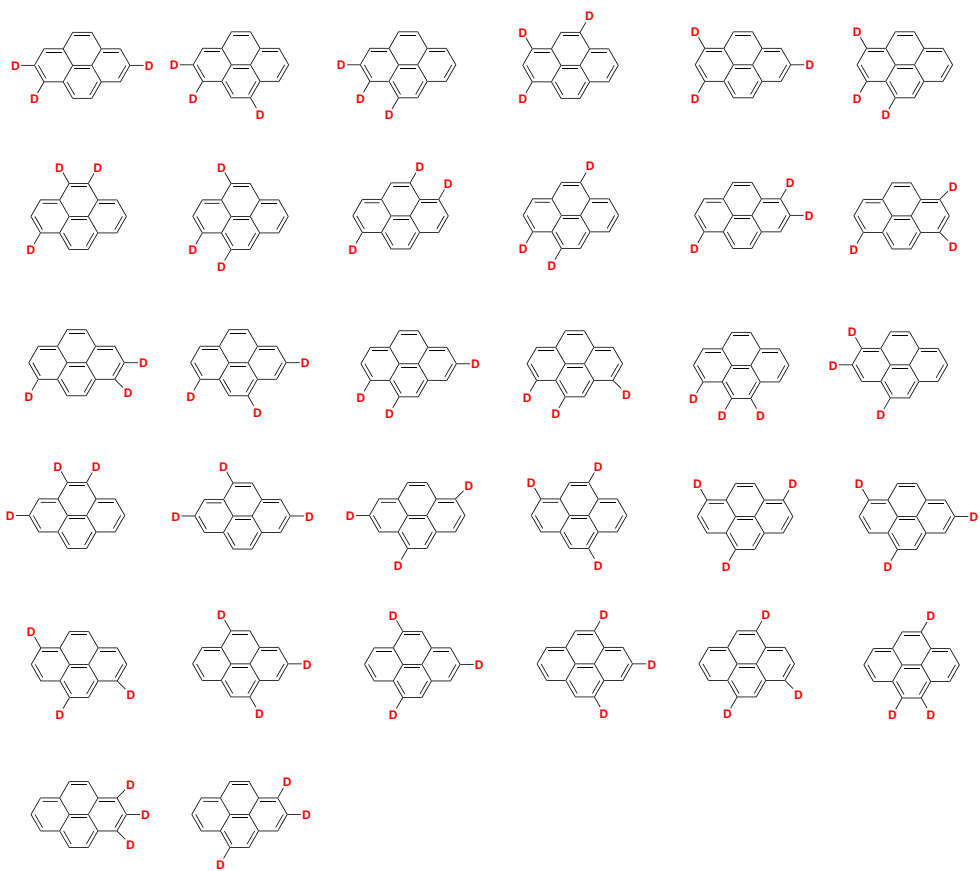}
}
\caption{\footnotesize
         \label{fig:Pyrene_3D_structures}
         Structures of the 32 isomers
         considered for triply-deuterated pyrenes
         (Pyre$\_$3D).
          }
\end{figure*}

\begin{figure*}
\centering{
\includegraphics[scale=0.5,clip]{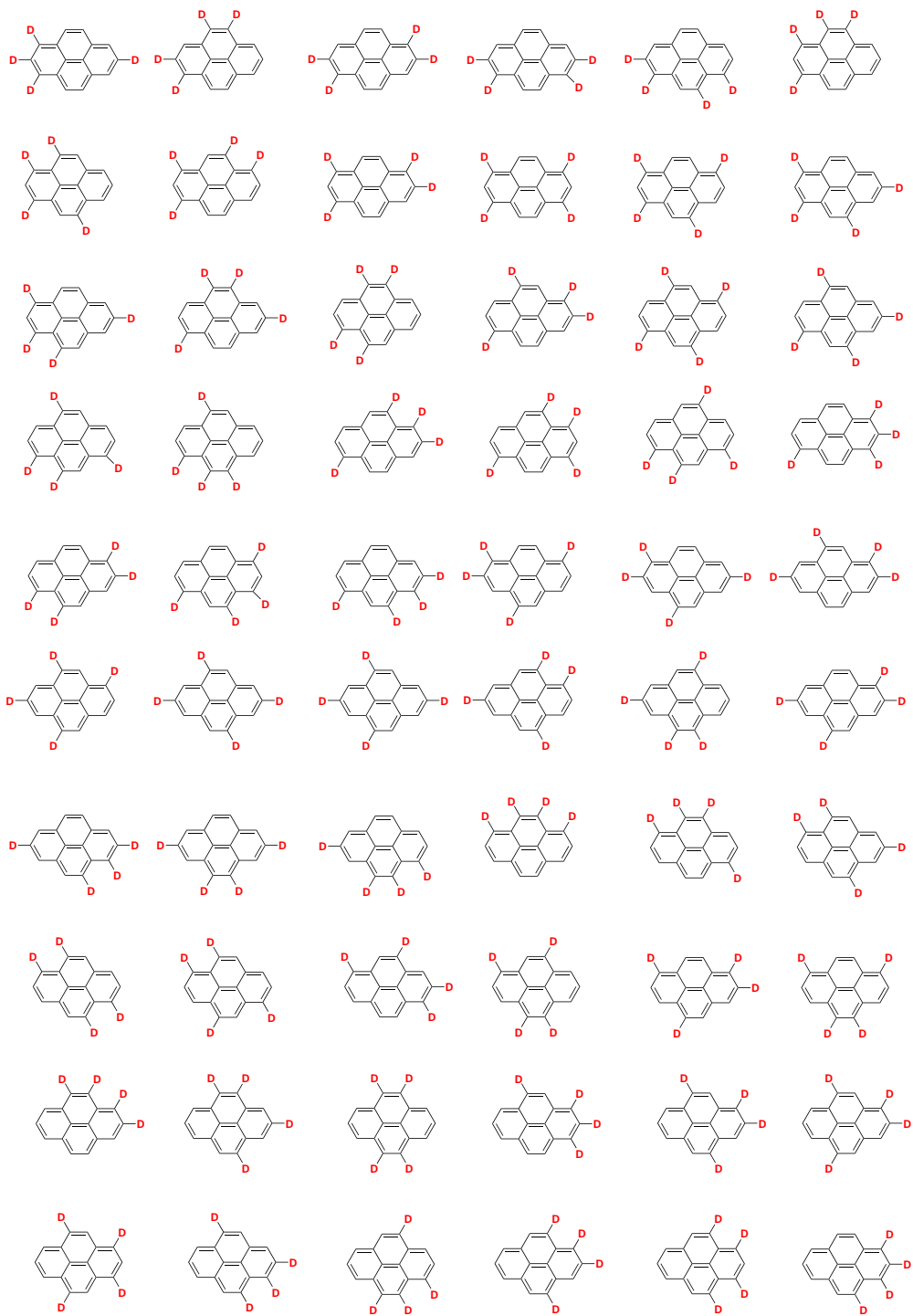}
}
\caption{\footnotesize
         \label{fig:Pyrene_4D_structures}
         Structures of the 60 isomers
         considered for quartet-deuterated pyrenes
         (Pyre$\_$4D).
         }
\end{figure*}

\begin{figure*}
\centering{
\includegraphics[scale=0.5,clip]{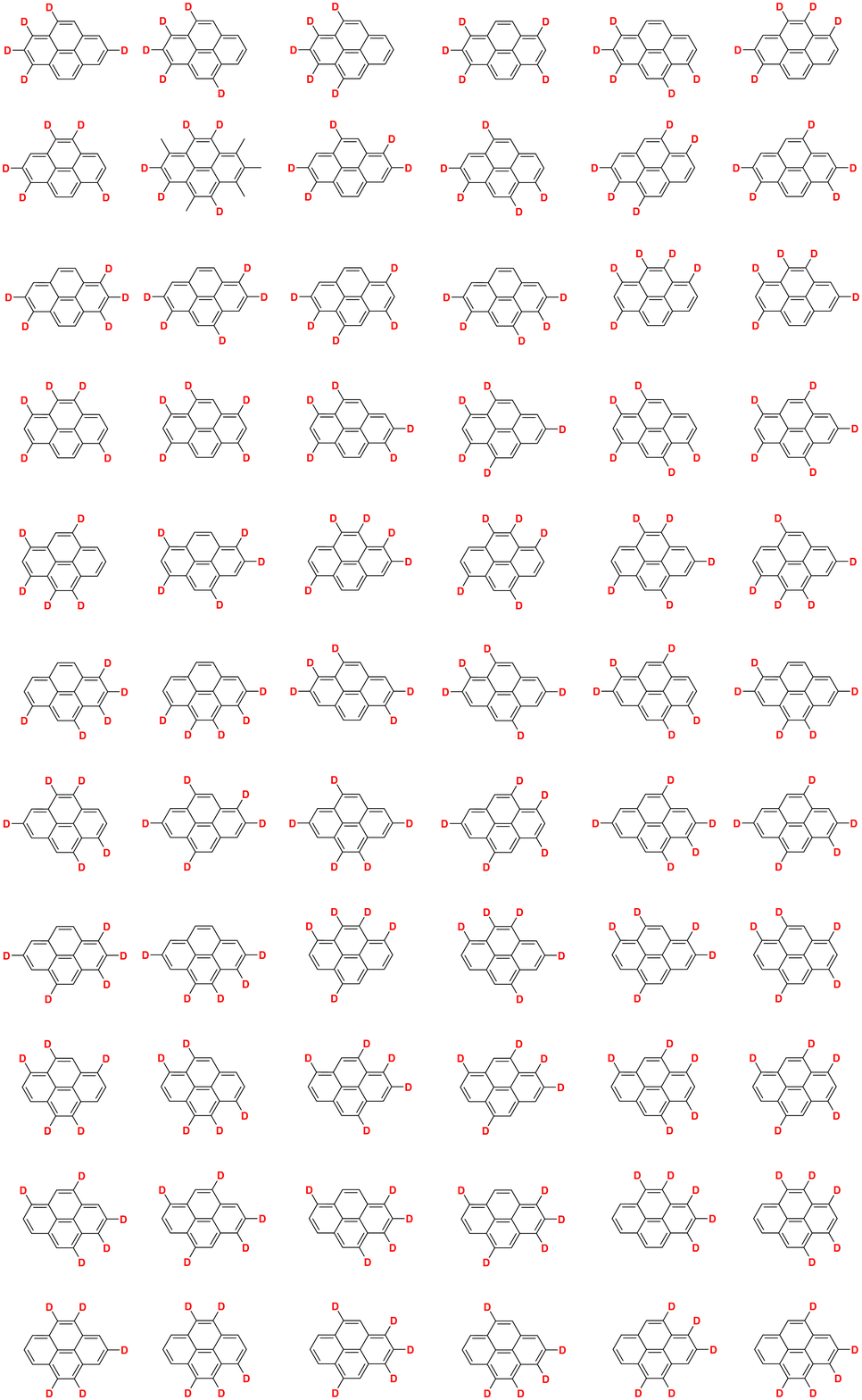}
}
\caption{\footnotesize
         \label{fig:Pyrene_5D_structures}
         Structures of the 66 isomers
         considered for quintet-deuterated pyrenes
         (Pyre$\_$5D).
         }
\end{figure*}

We first consider the multi-deuterated species
of pyrene which has a compact structure,
and is the smallest parent molecule in our sample.
We consider 288 isomers in total.
The isomers for mono-deuterated species
C$_{16}$H$_9$D (Pyre$\_$1D) and
di-deutereated species C$_{16}$H$_8$D$_2$
(Pyre$\_$2D) are shown in
Figure~\ref{fig:Pyrene_2D_structures},
while C$_{16}$H$_7$D$_3$ (Pyre$\_$3D),
C$_{16}$H$_6$D$_4$ (Pyre$\_$4D), and
C$_{16}$H$_5$D$_5$ (Pyre$\_$5D)
are displayed in
Figures~\ref{fig:Pyrene_3D_structures},
\ref{fig:Pyrene_4D_structures},
and \ref{fig:Pyrene_5D_structures},
respectively.
C$_{16}$H$_4$D$_6$ (Pyre$\_$6D),
C$_{16}$H$_3$D$_7$ (Pyre$\_$7D),
C$_{16}$H$_2$D$_8$ (Pyre$\_$8D), and
C$_{16}$H$_1$D$_9$ (Pyre$\_$9D)
are not shown since they are structurally
identical to Pyre$\_$4D, Pyre$\_$3D,
Pyre$\_$2D, and Pyre$\_$1D, respectively,
when exchanging H with D.

\begin{figure*}
\centering{
\includegraphics[scale=0.55,clip]{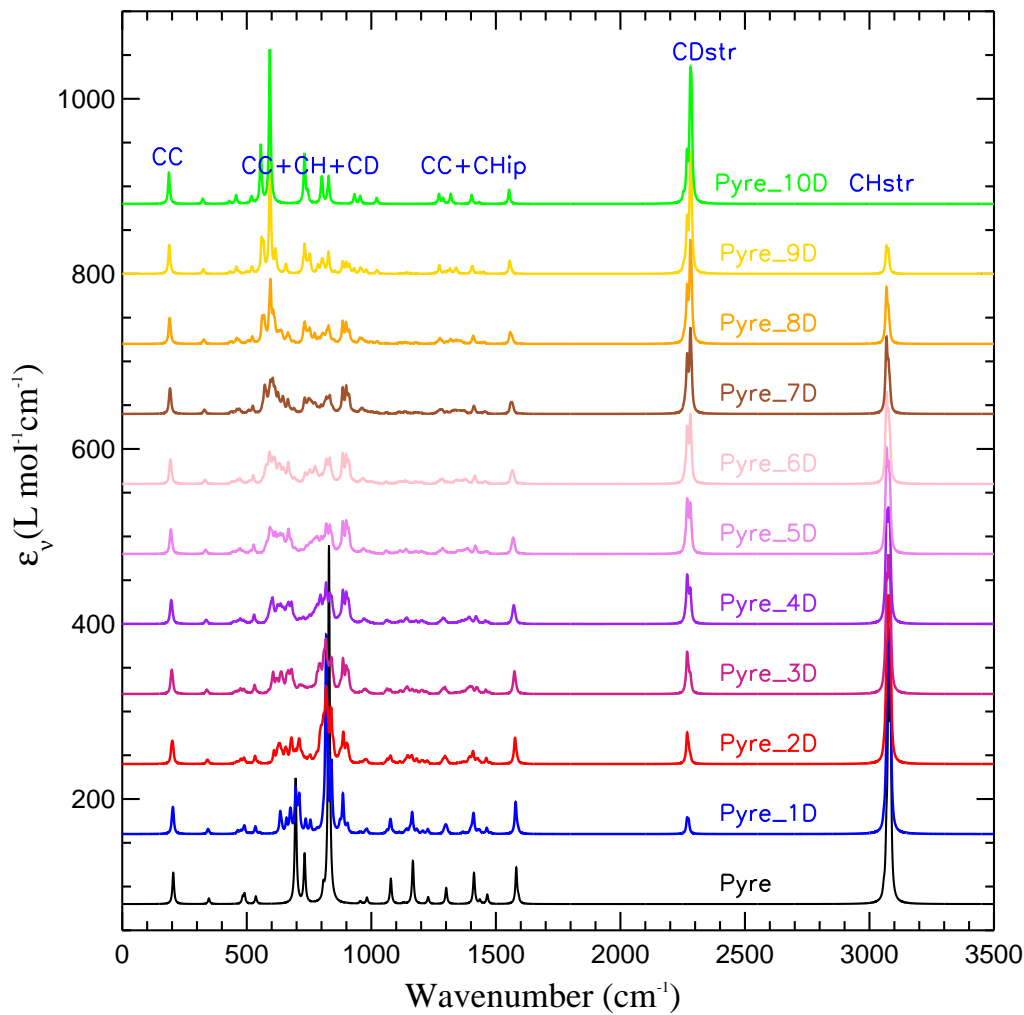}
}
\caption{\footnotesize
         \label{fig:Pyrene_nD_average_spec_all}
         DFT-computed vibrational absorption spectra
         of deuterated pyrenes of all possible deuterations,
         from zero-deuteration (i.e., pure pyrene C$_{16}$H$_{10}$),
         mono-deuteration (Pyre$\_$1D, C$_{16}$H$_{9}$D),
         di-deuteration (Pyre$\_$2D, C$_{16}$H$_{8}$D$_2$),
         all the way to perdeuteration
         (Pyre$\_$10D, C$_{16}$D$_{10}$).
         For each deuteration, the spectrum
         is obtained by averaging over
         all the considered isomers.
         The frequencies are scaled with
         a factor of 0.9688, and a line width
         of 4$\cm^{-1}$ is assigned.
         For perdeuterated pyrene, as expected,
         the 3.3$\mum$ C--H stretch is absent
         in the computed spectra
         due to the lack of H atoms in perdeuterated species.
         }
\end{figure*}

For each isomer we compute its vibrational
spectrum. We then obtain the mean spectrum
for deuterated pyrenes of a given degree of
deuteration (i.e., Pyre$\_$nD) by averaging
over all the isomers of  Pyre$\_$nD
(see Figure~\ref{fig:Pyrene_nD_average_spec_all}).
The frequencies are scaled with a factor of 0.9688
and each band is assigned with a line width of 4$\cm^{-1}$.
%

\begin{figure*}
\centering{
\includegraphics[scale=0.5,clip]{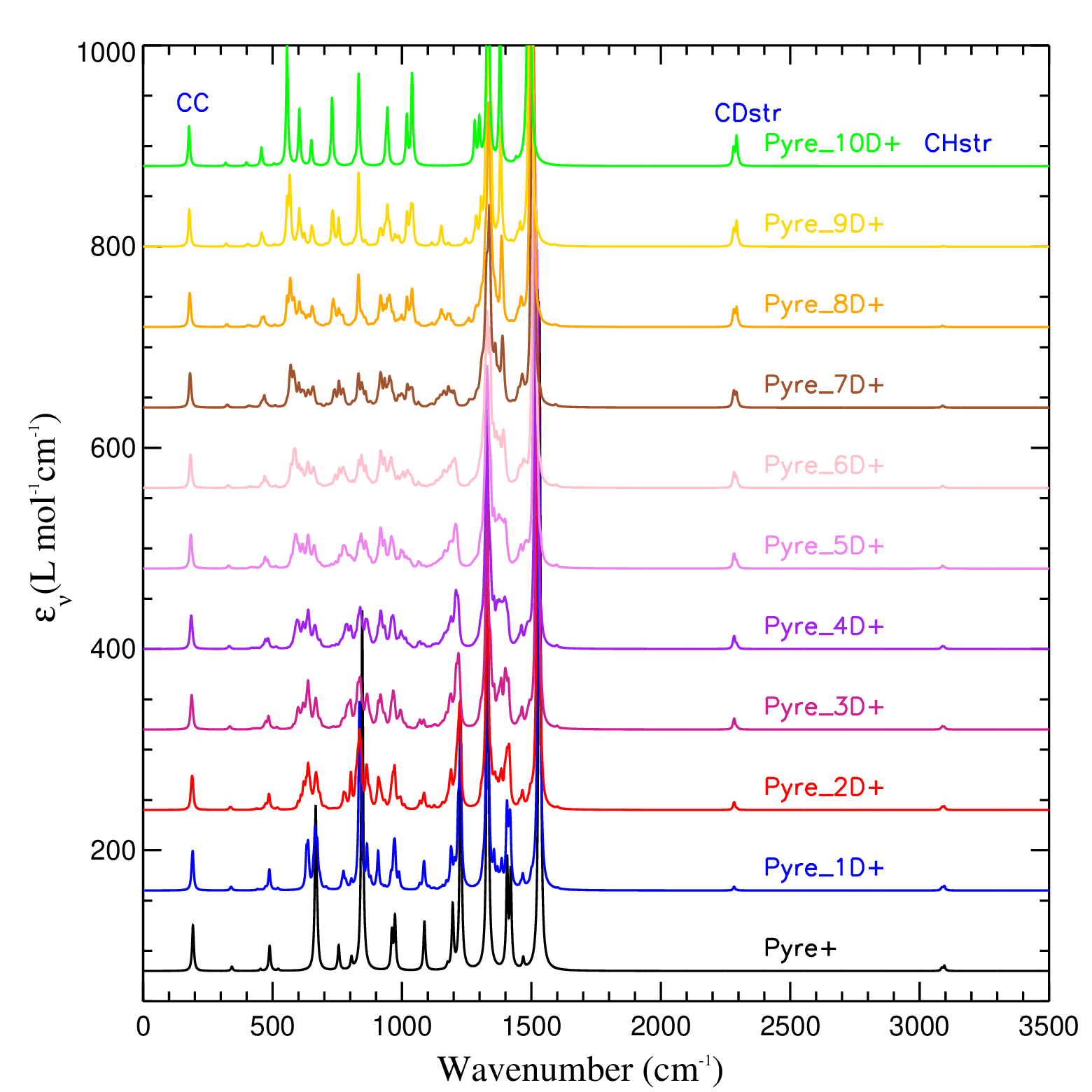}
}
\caption{\footnotesize
  \label{fig:Pyrene_nD_Plus_average_spec_all}
  Same as Figure~\ref{fig:Pyrene_nD_average_spec_all}
  but for deuterated pyrene cations.
         }
\end{figure*}

To explore the effects of ionization
on the spectra and band strengths of
deuterated pyrenes,
we also calculate the vibrational spectra
of all the 288 isomers for deuterated
pyrene cations (Pyre$\_$nD$^{+}$).
The mean spectrum for each deuteration
is shown in
Figure~\ref{fig:Pyrene_nD_Plus_average_spec_all}.

Figure~\ref{fig:Pyrene_nD_average_spec_all}
and Figure~\ref{fig:Pyrene_nD_Plus_average_spec_all}
clearly demonstrate that, upon deuteration,
the vibrational spectra of both neutral and
cationic pyrenes exhibit a new band
at $\simali$4.4$\mum$
or $\simali$2270$\cm^{-1}$
attributed to C--D stretch,
which is absent in the spectra
of pure pyrene and its cation.
The C--D in-plane ($\simali$11.7$\mum$)
and oop bending ($\simali$15.4$\mum$)
bands are also present
(and their frequencies and intensities
can be read from the Gaussian09 output files),
but difficult to identify
in Figures~\ref{fig:Pyrene_nD_average_spec_all},\ref{fig:Pyrene_nD_Plus_average_spec_all}
since they are mixed with
the C--H oop bending bands
and the C--C--C skeleton vibrational bands
(see Hudgins et al.\ 2004).

As expected,
Figure~\ref{fig:Pyrene_nD_average_spec_all}
and Figure~\ref{fig:Pyrene_nD_Plus_average_spec_all}
also clearly shows that,
as the deuteration increases,
the C--D stretch becomes stronger,
while the C--H stretch at 3.3$\mum$.
The latter is due to the reduction
in the number of C--H bonds
as pyrene becomes more deuterated.
%

By comparing
Figure~\ref{fig:Pyrene_nD_Plus_average_spec_all}
with Figure~\ref{fig:Pyrene_nD_average_spec_all},
one immediately sees that,
in comparison with their neutral counterparts,
in cations both the C--D and C--H stretches
are substantially suppressed
while the C--C stretches are considerably enhanced.
Such suppression in C--H stretches
and enhancement in C--C stretches
have also been seeen in pure PAHs
(e.g., see Allamandola et al.\ 1999),
mono-deuterated PAHs (Yang et al.\ 2020b),
superhydrogenated PAHs (Yang et al.\ 2020a)
and PAHs with aliphatic functional groups
(Yang et al.\ 2017).
Moreover, the C--D and C--H stretches
of cationic species appear to occur at
a somewhat shorter wavelength
compared with their neutral counterparts.

\begin{figure*}
\centering{
\includegraphics[scale=0.45,clip]{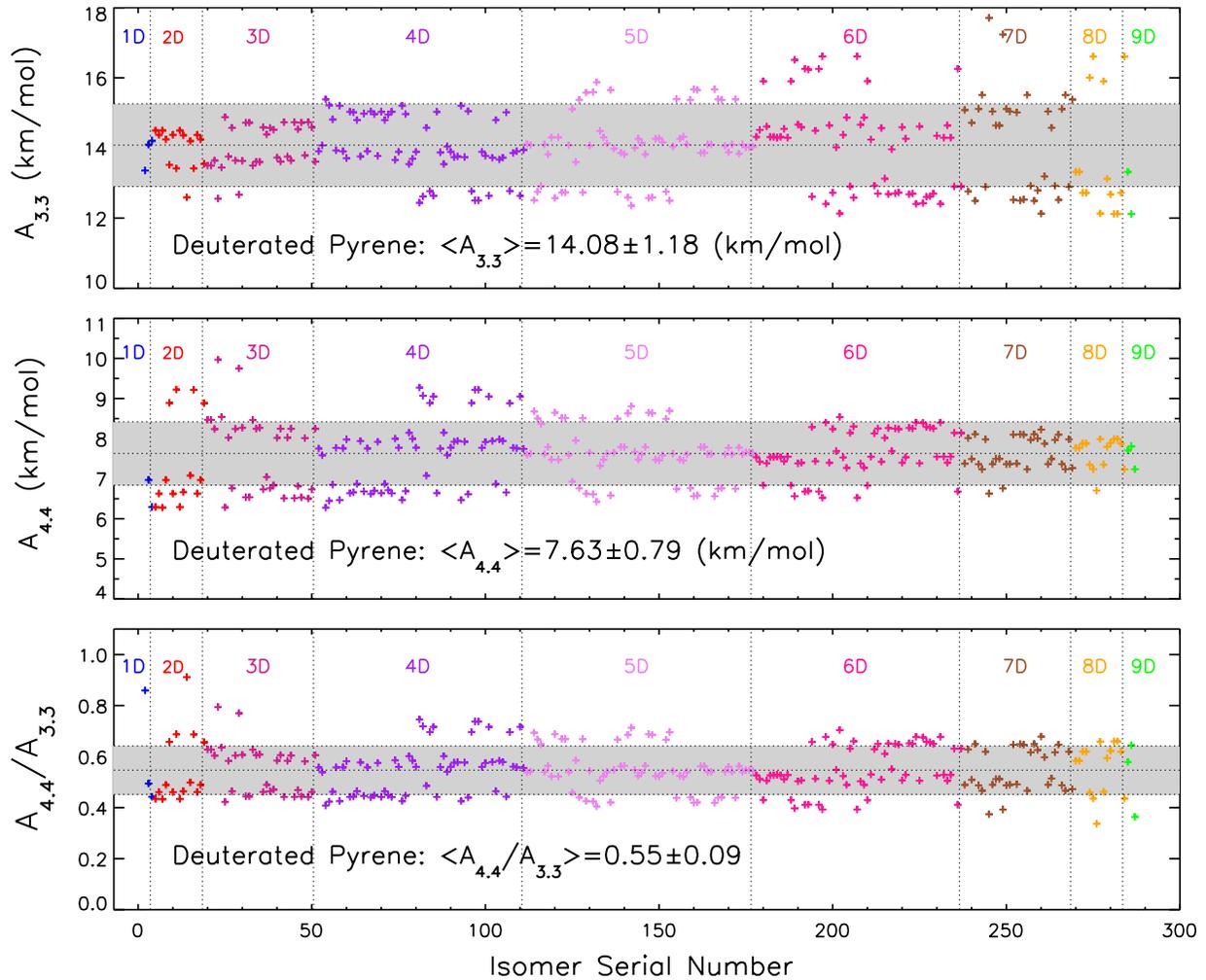}
}
\caption{\footnotesize
         \label{fig:Pyrene_nD_Aratio}
         Band strengths of the 3.3$\mum$ C--H stretches ($\ACH$)
         and the 4.4$\mum$ C--D stretches ($\ACD$)
         as well as the band-strength ratios $\ACD/\ACH$
         computed at level {\rm B3LYP/6-311+G$^{\ast\ast}$}
         for all the 288 isomers of deuterated pyrenes
         (Pyre$\_n$D) with different deuterations
         (i.e., $n$\,=\,1, 2, ..., 9).
         Perdeuterated ($n$\,=\,10) isomers
         are not shown since they do not have any
         C--H stretches.
         }
\end{figure*}

\begin{figure*}
\centering{
\includegraphics[scale=0.45,clip]{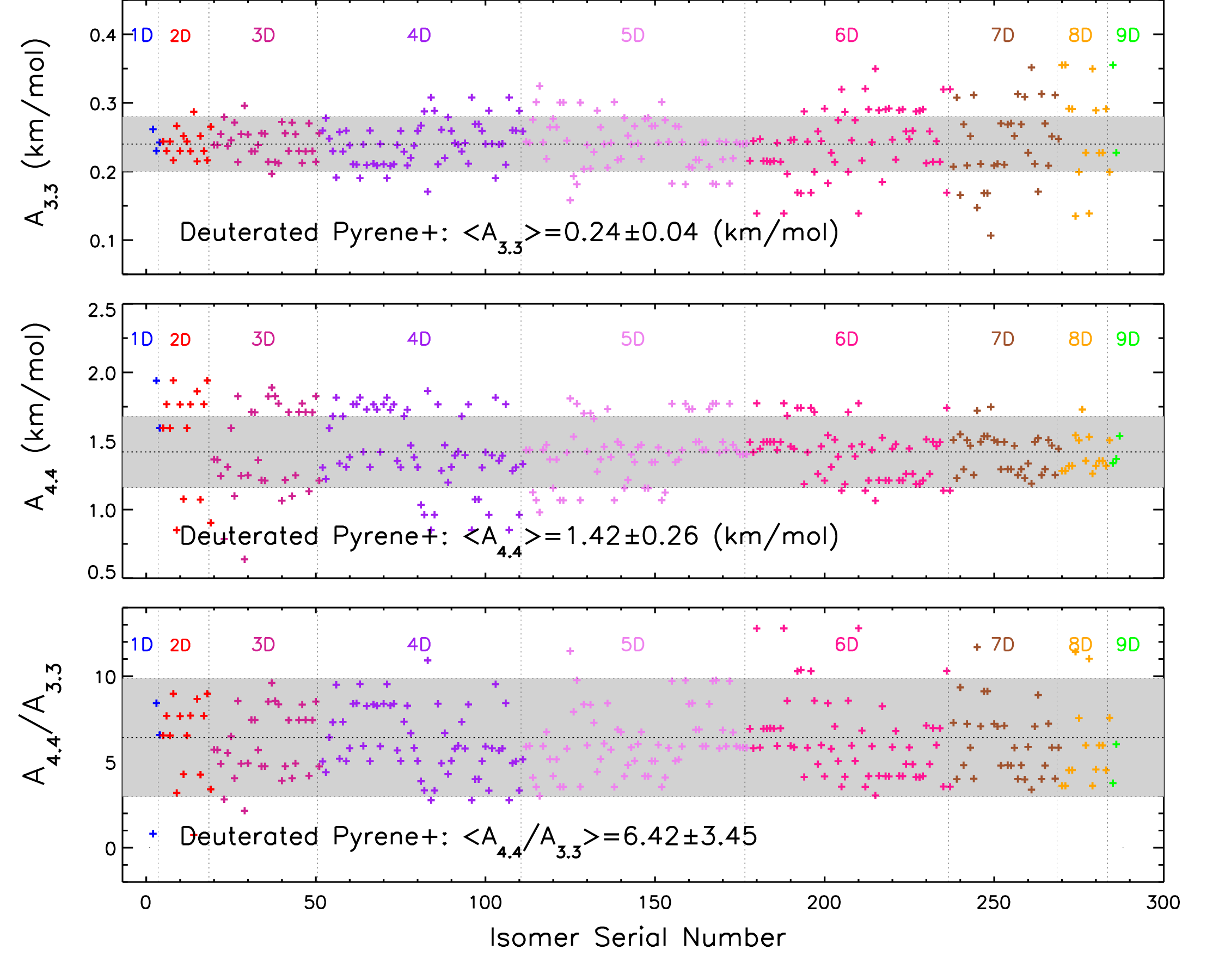}
}
\caption{\footnotesize
  \label{fig:Pyrene_nD_Plus_Aratio}
         Same as Figure~\ref{fig:Pyrene_nD_Aratio}
         but for deuterated pyrene cations.
         }
\end{figure*}

As mentioned earlier in this section,
while the C--D in-plane and
oop bending bands of multi-deuterated
pyrenes often blend with other bands,
the C--D and C--H stretches lie in
such a ``clean'' spectral region that we
can unambiguously determine their
frequencies and intensities.
In the following we will
focus on the 3.3$\mum$ C--H stretch
and the 4.4$\mum$ C--D stretch.
We obtain from DFT computations
the intensities of the C--H and C--D stretches
for all the 288 isomers of deuterated pyrenes.
Let $\Aaro$ and $\Acd$ respectively
be the intrinsic band strengths of
the aromatic C--H and C--D stretches
on a per bond basis.
We show in Figure~\ref{fig:Pyrene_nD_Aratio}
and Figure~\ref{fig:Pyrene_nD_Plus_Aratio}
the intensities ($\Aaro$, $\Acd$)
and their ratios (i.e., $\Acd/\Aaro$)
for all the 288 isomers of deuterated pyrenes
and their cations, respectively.
In Tables~\ref{tab:Freq_Int_Pyre_nD_all},\ref{tab:Freq_Int_Pyre_nDPlus_all}
we tabulate the {\it mean} wavelengths and
intensities of the C--H and C--D stretches
for each deuteration $n$
(i.e., Pyre$\_n$D and Pyre$\_n$D$^+$,
with $n$\,=\,0, 1, 2, ..., 10),
obtained by averaging over all the isomers
of a given deuteration $n$.
%
%

The standard deviations for the mean wavelengths
of the C--H and C--D stretches are very small
for both neutral and cationic species,
indicating that the central wavelengths of
these two vibrational modes are very similar
for different deuterations.
By averaging over all deuterations,
we obtain the mean band strengths
$\langle\ACH\rangle\approx14.08\pm1.18\km\mol^{-1}$,
$\langle\ACD\rangle\approx7.63\pm0.79\km\mol^{-1}$,
and $\langle\ACD/\ACH
\rangle\approx0.55\pm0.09$
for neutral pyrenes,
and $\langle\ACH\rangle\approx0.24\pm0.04\km\mol^{-1}$,
$\langle\ACD\rangle\approx1.42\pm0.26\km\mol^{-1}$,
and $\langle\ACD/\ACH\rangle\approx6.42\pm3.45$
for cationic pyrenes.
It is interesting to note that, overall,
the band strengths of the C--H and C--D stretches
of multi-deuterated pyrenes and their cations vary
little with deuteration, although the band ratios
$\langle\ACD/\ACH\rangle$ of pyrene cations
exhibit a relatively larger scatter.
%

\subsection{Multi-Deuterated Perylene, Coronene and Ovalene
  \label{sec:largerPAHs}}
To explore the effects of PAH size on
the band strengths of the C--H and C--D stretches,
we now consider the multi-deuterated
derivatives of three compact molecules
which all are larger than the four-ring pyrene
(see \S\ref{sec:pyrene}):
the 5-ring perylene,
the 7-ring coronene, and
the 10-ring ovalene
(see Figure~\ref{fig:PAH-MultiD_structures}).
Multi-deuterated perylene and ovalene
have a huge number of isomers and it is
impractical to compute the vibrational spectra
of all their isomers.
Therefore, 1072 isomers of perylene
and 2462 isomers of ovalene are randomly
selected for DFT computations.
For coronene, we consider all of the 382 isomers
of its multi-deuterated derivatives
since it is highly symmetric and do not
have as many ``inequivalent'' isomers
as perylene and ovalene.
As demonstrated in \S\ref{sec:pyrene}
and in Yang et al.\ (2020b),
the 4.4$\mum$ C--D stretch is substantially
suppressed in deuterated PAH cations,
in the following we will only consider
neutral species since the 4.4$\mum$
C--D band seen in astrophysical regions is
expected to predominantly arise from
deuterated neutral PAHs.

\begin{figure*}
\centering{
\includegraphics[scale=0.5,clip]{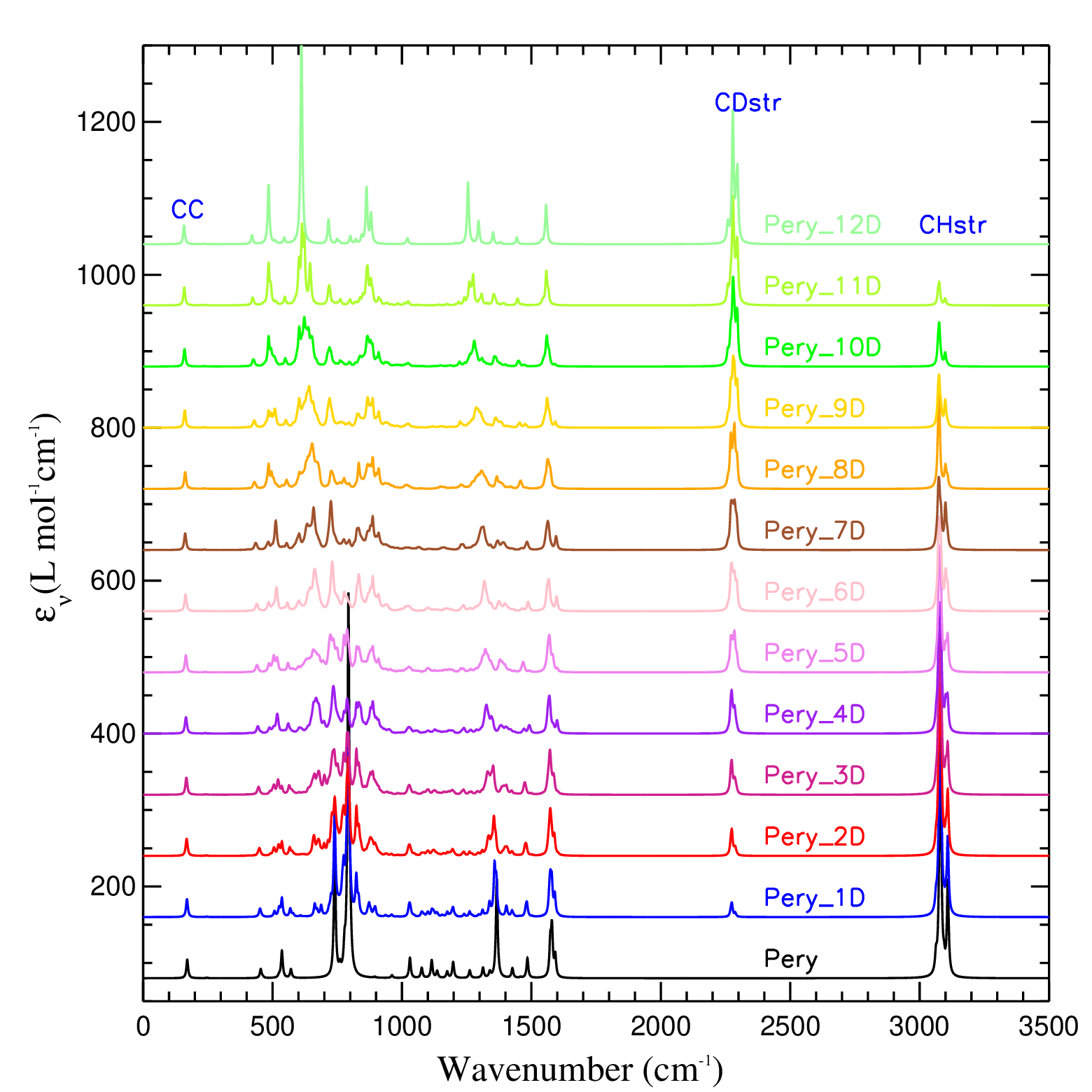}
}
\caption{\footnotesize
  \label{fig:Perylene_nD_average_spec_all}
  Same as Figure~\ref{fig:Pyrene_nD_average_spec_all}
  but for deuterated perylenes.
         }
\end{figure*}

\begin{figure*}
\centering{
\includegraphics[scale=0.5,clip]{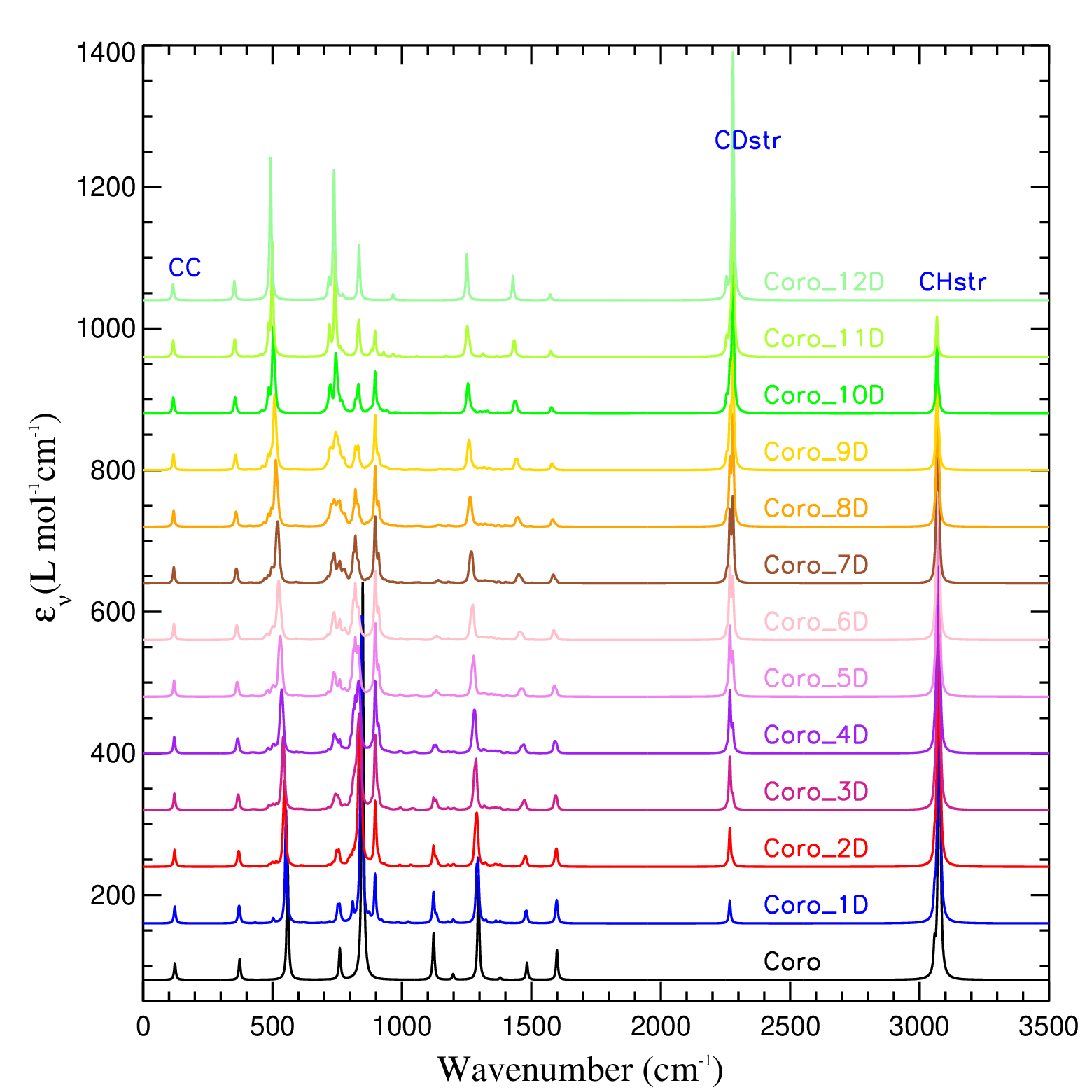}
}
\caption{\footnotesize
  \label{fig:Coronene_nD_average_spec_all}
  Same as Figure~\ref{fig:Pyrene_nD_average_spec_all}
  but for deuterated coronenes.
           }
\end{figure*}

\begin{figure*}
\centering{
\includegraphics[scale=0.5,clip]{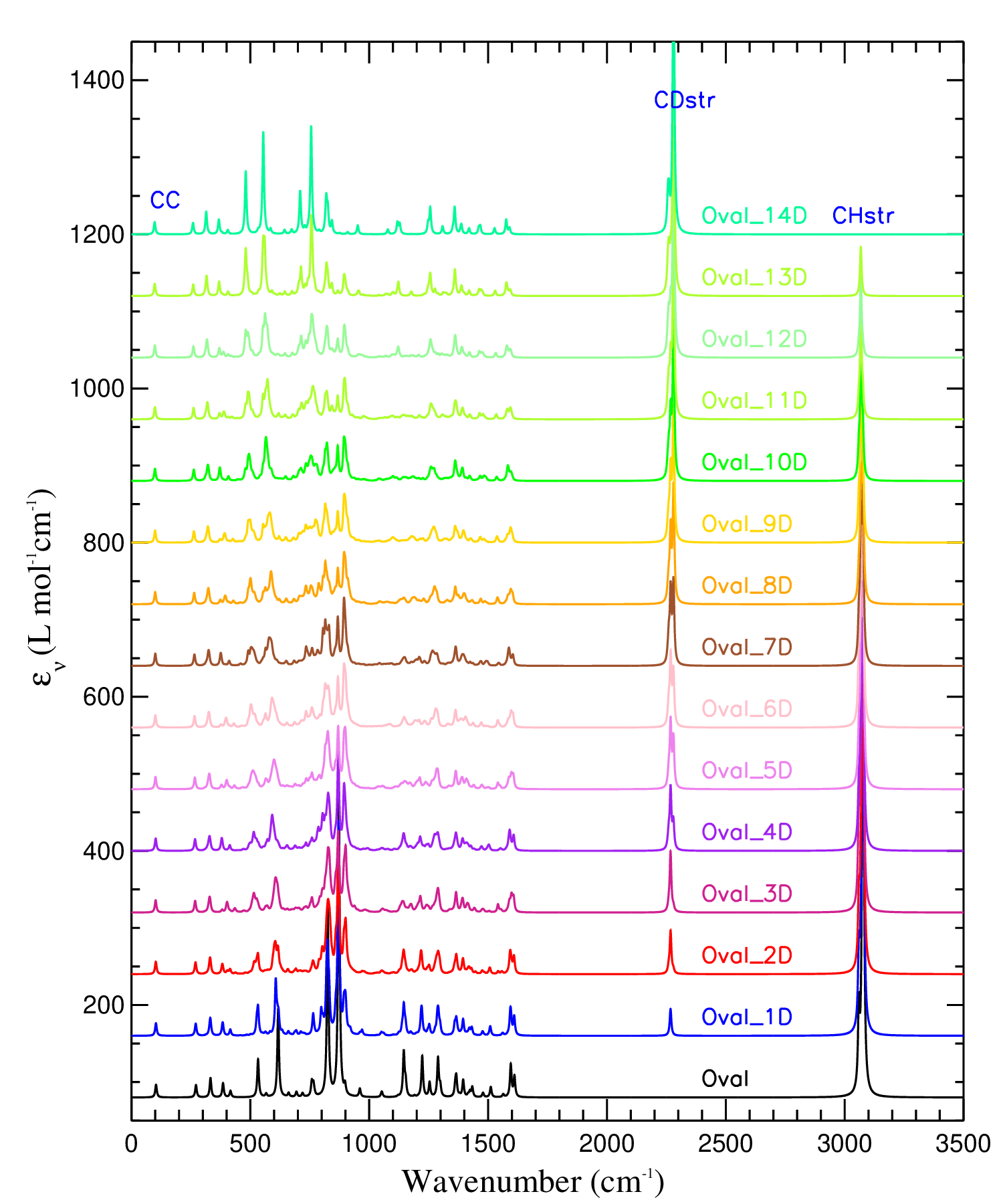}
}
\caption{\footnotesize
  \label{fig:Ovalene_nD_average_spec_all}
  Same as Figure~\ref{fig:Pyrene_nD_average_spec_all}
  but for deuterated ovalenes.
         }
\end{figure*}

In
Figures~\ref{fig:Perylene_nD_average_spec_all}--\ref{fig:Ovalene_nD_average_spec_all}
we respectively show the mean vibrational spectra
of a given deuteration $n$ for multi-deuterated
perylenes (Pery$\_n$D), coronenes (Coro$\_n$D),
and ovalenes (Oval$\_n$D).
Just like that of multi-deuterated pyrenes,
the C--D stretch is clearly seen
at $\simali$4.4$\mum$
($\simali$2250$\cm^{-1}$)
in the computed spectra of all deuterated species.
Also, the C--D stretch becomes appreciably stronger
while the C--H strecth at 3.3$\mum$
($\simali$3000$\cm^{-1}$) weakens
as the deuteration increases.
Again, the C--D in-plane ($\simali$850$\cm^{-1}$)
and oop bending modes ($\simali$650$\cm^{-1}$)
are well mixed with the C--H oop bending modes
and the C--C--C skeleton vibrational modes,
and therefore they cannot be clearly identified
in the mean spectra.
Meanwhile, the intensities of the weak bands
at $\lambda^{-1}\simlt 400\cm^{-1}$,
which arise from the C--C--C skeleton vibrations,
are basically unaffected by deuteration,
while their frequencies tend to be redshifted
as deuteration increases.

\begin{figure*}
\centering{
\includegraphics[scale=0.45,clip]{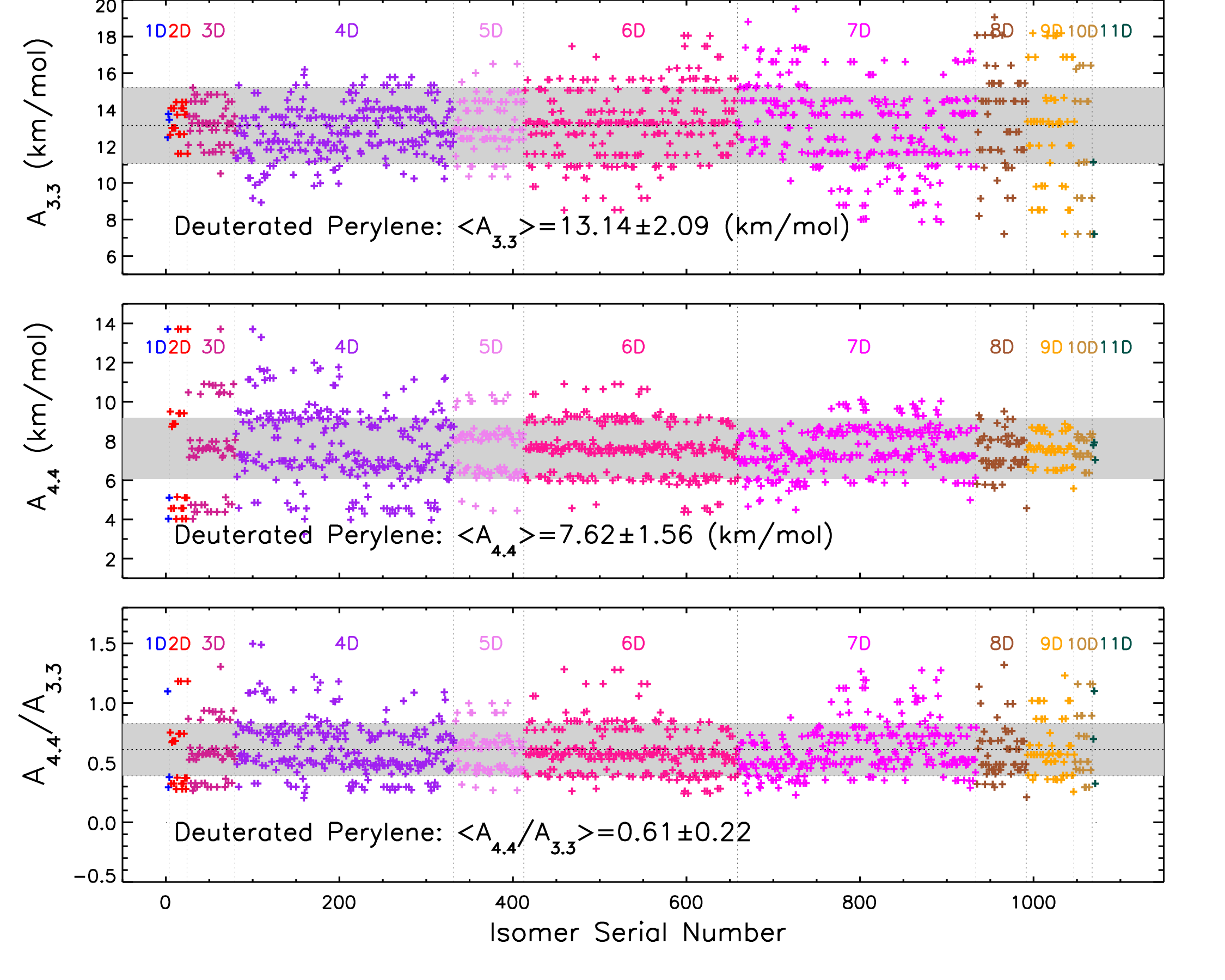}
}
\caption{\footnotesize
  \label{fig:Perylene_nD_Aratio}
   Same as Figure~\ref{fig:Pyrene_nD_Aratio}
   but for deuterated perylenes.
           }
\end{figure*}

\begin{figure*}
\centering{
\includegraphics[scale=0.45,clip]{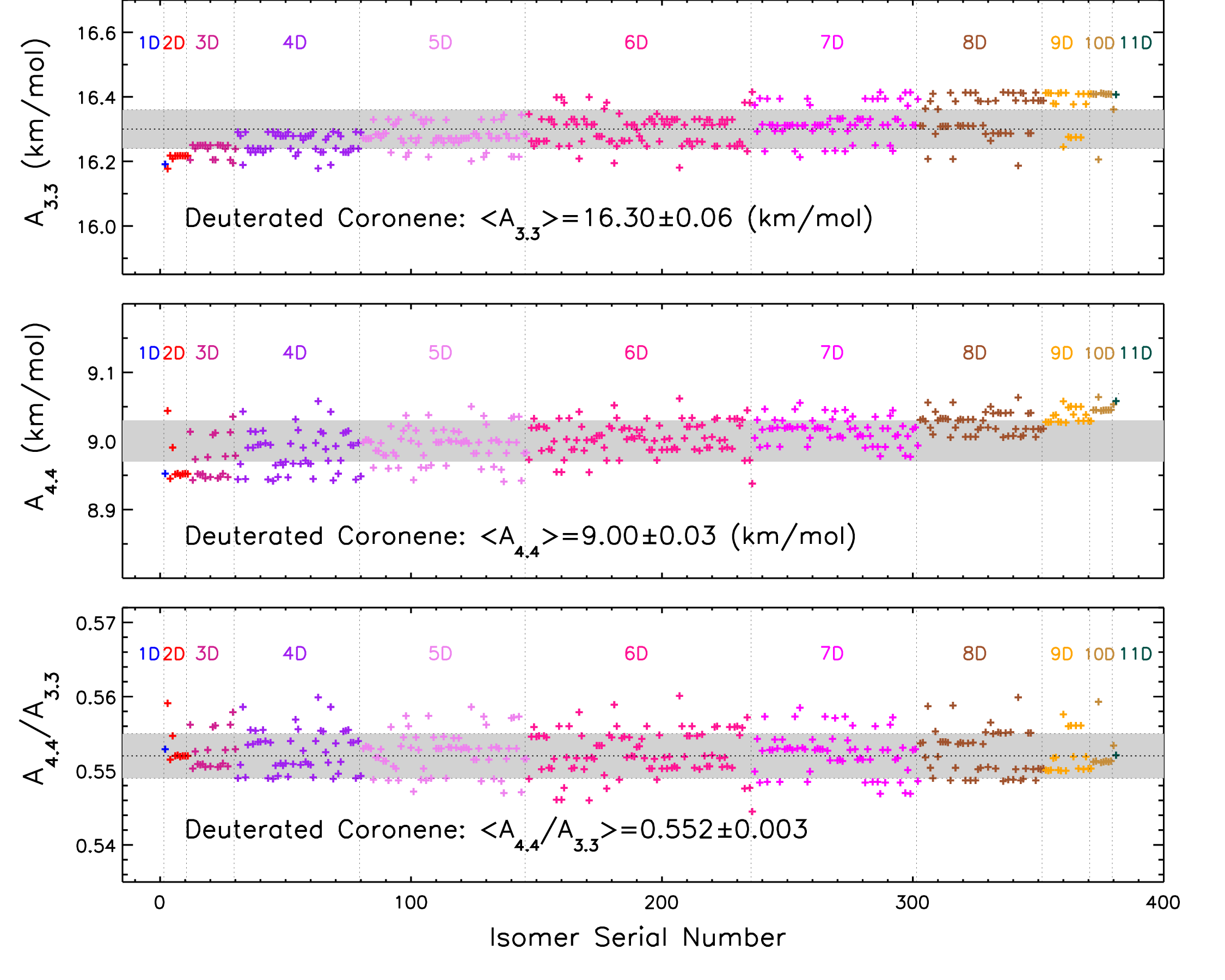}
}
\caption{\footnotesize
  \label{fig:Coronene_nD_Aratio}
  Same as Figure~\ref{fig:Pyrene_nD_Aratio}
   but for deuterated coronenes.
         }
\end{figure*}

\begin{figure*}
\centering{
\includegraphics[scale=0.4,clip]{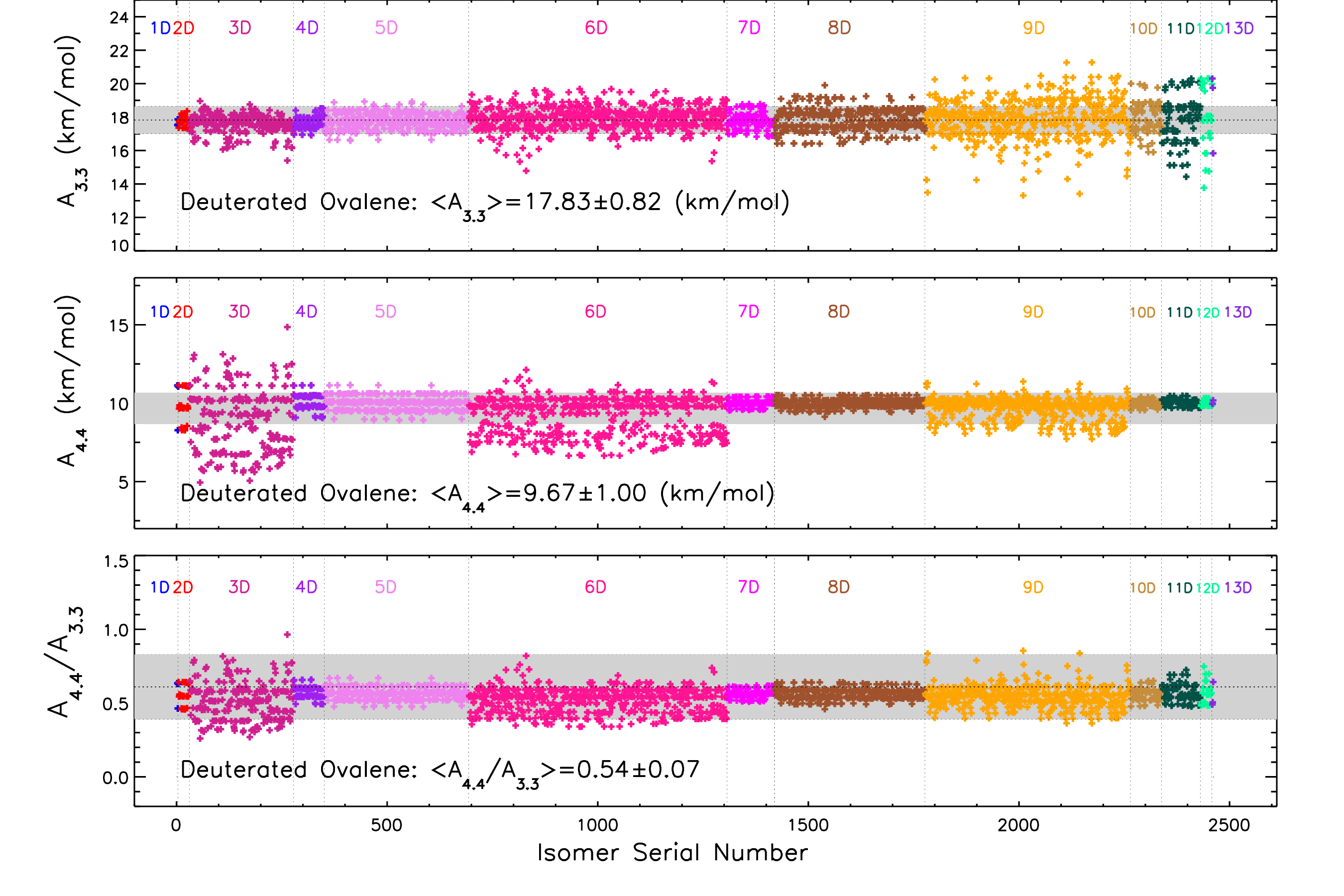}
}
\caption{\footnotesize
  \label{fig:Ovalene_nD_Aratio}
  Same as Figure~\ref{fig:Pyrene_nD_Aratio}
   but for deuterated ovalenes.
         }
\end{figure*}

In Figures~\ref{fig:Perylene_nD_Aratio}--\ref{fig:Ovalene_nD_Aratio}
we show the intensities of the C--H stretch ($\ACH$)
and the C--D stretch ($\ACD$)
and their ratios ($\Aratio$)
respectively for all the isomers of
Pery$\_n$D, Coro$\_n$D,
and Oval$\_n$D of all possible deuterations
(i.e., $n$ ranges from 1 all the way to
12 for perylene, 12 for coronene,
and 14 for ovalene).
The overall mean wavelengths and intensities
of the C--H and C--D stretches obtained by
averaging over all the deuterations,
as well as those of each deuteration $n$
are tabulated in
Tables~\ref{tab:Freq_Int_Pery_nD_all}--\ref{tab:Freq_Int_Oval_nD_all}
for Pery$\_$nD, Coro$\_$nD, and Oval$\_$nD, respectively.
It is apparent that both the wavelengths
and the intensities of the C--H and
C--D stretches (on a per unit bond basis)
are closely similar among deuterated species
of different deuterations.
While the intensities of the C--H and C--D
stretches of Coro$\_$nD and Oval$\_$nD
seem to exceed that of pyrene and perylene
(as well as tetracene, see \S\ref{sec:tetracene})
by $\simali$30\%,
their $\Aratio$ ratios are rather similar
to that of the deuterated derivatives of
pyrene, perylene and tetracene.

\subsection{Multi-Deuterated Tetracene\label{sec:tetracene}}
In previous sections, the molecules considered
so far (i.e., pyrene, perylene, coronene, and ovalene)
are all compact and pericondensed.
To explore the effects of the molecular structures
or shapes on the intrinsic strengths of the C--H
and C--D stretches, we consider tetracene
(C$_{18}$H$_{12}$), a linear, catacondensed
molecule (see Figure~\ref{fig:PAH-MultiD_structures}),
as an extreme case, even though catacondensed
molecules are less likely to survive in the hostile ISM.

\begin{figure*}
\centering{
\includegraphics[scale=0.5,clip]{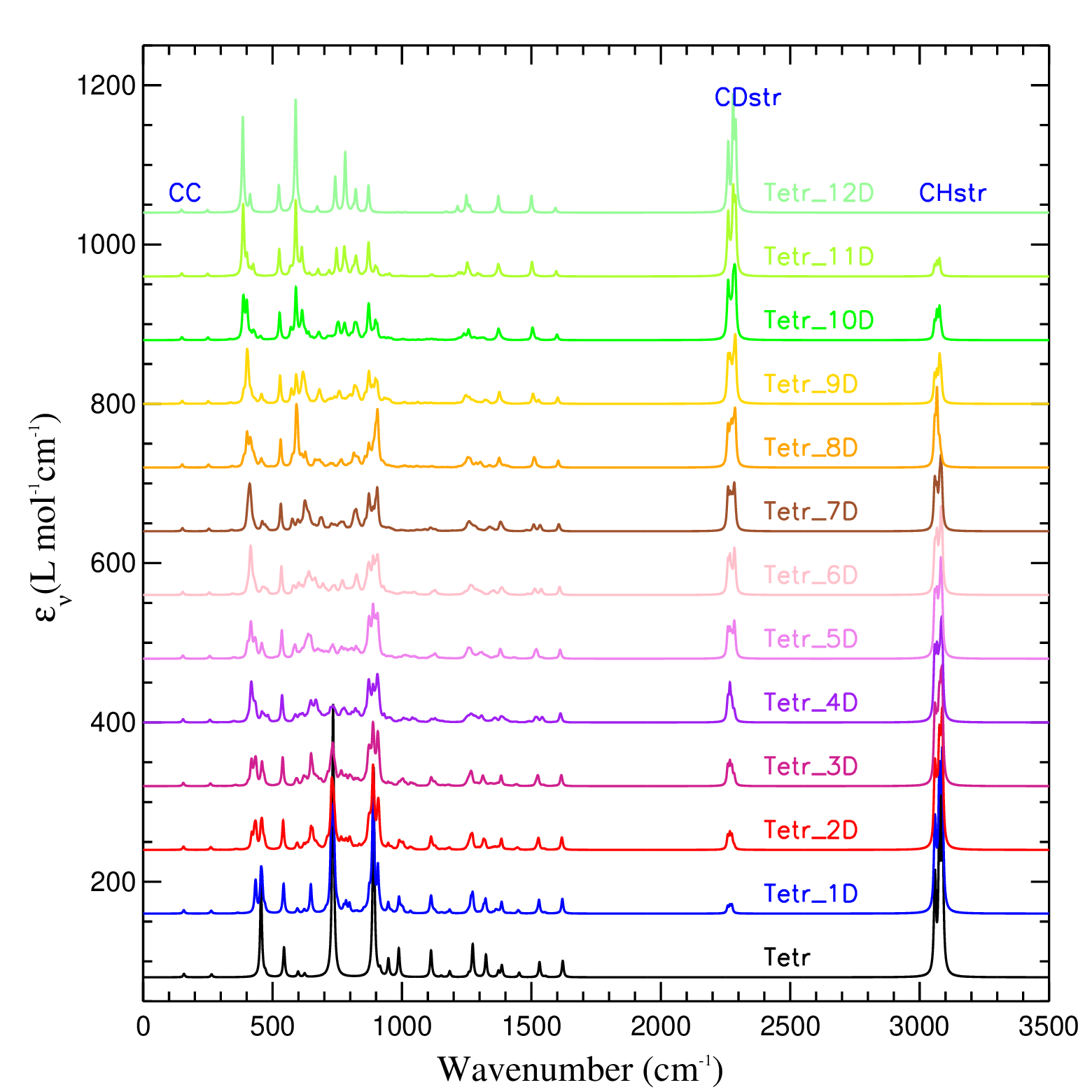}
}
\caption{\footnotesize
  \label{fig:Tetracene_nD_average_spec_all}
  Same as Figure~\ref{fig:Pyrene_nD_average_spec_all}
  but for deuterated tetracenes.
         }
\end{figure*}

\begin{figure*}
\centering{
\includegraphics[scale=0.45,clip]{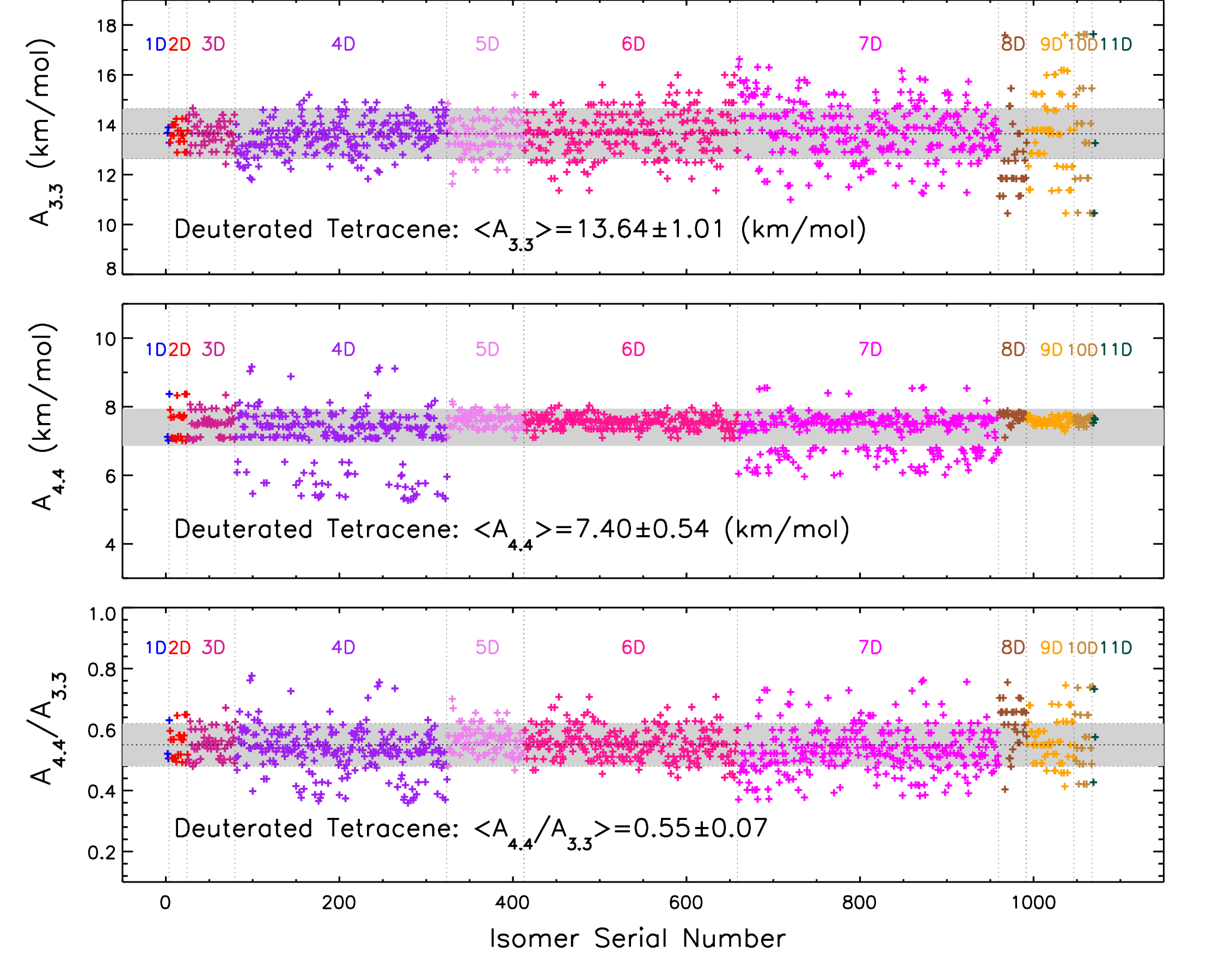}
}
\caption{\footnotesize
  \label{fig:Tetracene_nD_Aratio}
   Same as Figure~\ref{fig:Pyrene_nD_Aratio}
   but for deuterated tetracenes.
         }
\end{figure*}

Similar to multi-deuterated perylenes and ovalenes,
multi-deuterated tetracenes have too many isomers
to be computationally manageable.
Therefore, we randomly select 1072 isomers
and compute their vibrational spectra.
In Figure~\ref{fig:Tetracene_nD_average_spec_all}
we show the mean spectra of all the 12 deuterations
of tetracene (i.e., Tetr$\_n$D, where $n$\,=\,1, 2, ..., 12).
Again, the 4.4$\mum$ C--D stretch is present
in the computed spectra of all the isomers
and becomes more pronounced as the molecule
becomes more deuterated.
It is worth noting that, compared with
that of the deuterated derivatives of
compact species (i.e., pyrene, perylene,
coronene, and ovalene),
the C--C--C skeleton vibrations
at $\lambda^{-1}\simlt 400\cm^{-1}$
are considerably suppressed in deuterated tetracenes.
The mean intensities of the C--H and C--D stretches
obtained by averaging over all the isomers of a given
deuteration $n$ are tabulated in
Table~\ref{tab:Freq_Int_Tetr_nD_all}.
Also tabulated are the overall mean intensities
averaged over all the deuterations.
Figure~\ref{fig:Tetracene_nD_Aratio}
shows the intrinsic band strengths of
the C--H and C--D stretches and their ratios
($\Aratio$) computed for all the 1072 isomers
of deuterated tetracenes.
It is apparent that both the mean band strengths
$\ACH$ and $\ACD$ as well as $\Aratio$ of
deuterated tetracenes do not vary much with
deuteration and are closely consistent with
that of multi-deuterated pyrenes and perylenes
as well as mono-deuterated species
(see Yang et al.\ 2020b).
This demonstrates that the intrinsic
band strengths $\ACH$ and $\ACD$
and their ratios $\Aratio$ of deuterated
species do not seem to exhibit any appreciable
variations with the molecular structures of PAHs.

\section{Discussion and Astrophysical
            Implications}\label{sec:astro}
As shown in \S\ref{sec:pyrene},
\S\ref{sec:largerPAHs} and \S\ref{sec:tetracene},
the intrinsic band strengths of the C--H ($\ACH$)
and C--D ($\ACD$) stretches and particularly their
ratios ($\Aratio$) vary very little with the degree of
deuteration, and the size and structure
of the parent molecule,
just as expected from the BO approximation,
one of the foundations of the B3LYP DFT
(see \S\ref{subsec:BO}).
Also, the wavelengths of the C--H and C--D
stretches peak at $3.25\pm0.1\mum$
and $4.40\pm0.1\mum$, respectively,
and exhibit very little variations with
the degree of deuteration, and the size
and structure of the parent molecule
(see Tables~\ref{tab:Freq_Int_Pyre_nD_all}--\ref{tab:Freq_Int_Tetr_nD_all}),
which is also expected from the BO approximation.
As a matter of fact, previous DFT calculations
on the vibrational spectra of pure PAHs of various sizes
and structures (e.g., see Bauschlicher et al.\ 2008,
2009, 2018; Mattioda et al.\ 2020)
have shown that the wavelengths and intensities
of the C--H stretches are insensitive to
the PAH size and structure.
This is also true for the aromatic C--H stretches
of PAHs attached with aliphatic sidegroups
(e.g., see Yang et al.\ 2013, 2016).
These can all be understood in the context
of the BO approximation. 
%
%
%
%
%

We should note that, as shown in
Tables~\ref{tab:Freq_Int_Pyre_nD_all}--\ref{tab:Freq_Int_Tetr_nD_all},
the C--H stretches of deuterated PAHs peak
at $\simali$3.25$\mum$ which is appreciably
shorter than the nominal ``3.3$\mum$'' UIE band
which typically occurs at $\simali$3.28--3.30$\mum$.
Such a wavelength mismatch had been long noted
by Sakata et al.\ (1990) and Kwok \& Zhang (2013)
for {\it pure} PAH molecules of which the C--H stretches
lie shortward of the observed wavelength.
Perhaps the wavelength mismatch could be accounted
for by PAHs including substituents
(e.g., N in place of C; see Hudgins et al.\ 2005,
Mattioda et al.\ 2008)
or superhydrogenation (Bernstein et al.\ 1996,
Thrower et al.\ 2012, Sandford et al.\ 2013,
Yang et al.\ 2020a, Cruz-Diaz et al.\ 2020).
It is well recognized that upon substitution,
the central wavelengths of the C--C and C--H
bands of PAHs appreciably shift. Indeed,
the observed subtle variations in the peak
wavelength of  the 6.2$\mum$ C--C stretching
emission band were commonly attributed to
polycyclic aromatic nitrogen heterocycles---PAHs
with one or more nitrogen atoms substituted into
their carbon skeleton (see Hudgins et al.\ 2005).
Also, both experimental measurements
and DFT computations have shown that 
in superhydrogenated PAHs the wavelength
of the C--H stretch shifts to longer wavelengths
(see Sandford et al.\ 2013, Yang et al.\ 2020a).
%

%
By averaging the band strengths
(on a per unit bond basis)
over neutral PAHs of different deuteration,
size, and structure, we obtain a recommended,
mean band strength of
$\langle\Aaro\rangle\approx 15.3\pm2.1\km\mol^{-1}$
for the 3.3$\mum$ C--H stretch,
$\langle\ACD\rangle\approx 8.5\pm1.0\km\mol^{-1}$
for the 4.4$\mum$ C--D stretch, and
a band-strength ratio of
$\langle\ACD/\Aaro\rangle\approx0.56\pm0.03$,
which are closely similar to that of mono-deuterated PAHs
($\langle\Aaro\rangle\approx13.2\pm1.0\km\mol^{-1}$,
$\langle\Aaro\rangle\approx7.3\pm2.4\km\mol^{-1}$,
and $\langle\ACD/\Aaro\rangle\approx0.56\pm0.19$).

Following Yang et al.\ (2020b),
we define the degree of deuteration of
a deuterated PAH molecule consisting of
$\NH$ H atoms and $\ND$ D atoms
as $\Dpah\equiv\ND/\left(\NH+\ND\right)$.
Yang et al.\ (2020b) showed that one can
derive $\Dpah$ from
\begin{equation}
\label{eq:NDNH}
 \Dpah \approx
\left\{
1\,+\,\left(\frac{I_{3.3}}{I_{4.4}}\right)_{\rm obs}
\times\left(\frac{\Acd}{\Aaro}\right)
\times \left(\frac{B_{4.4}}{B_{3.3}}\right)
\right\}^{-1} ~~,
\end{equation}
where $\Iratioobs$ is the {\it observed} ratio
of the power emitted from the 3.3$\mum$
aromatic C--H band ($I_{3.3}$) to that from
the 4.4$\mum$ aromatic C--D band ($I_{4.4}$),
$B_{3.3}$ and $B_{4.4}$ are respectively
the Planck functions at  temperature $T$
and wavelengths 3.3$\mum$ and 4.4$\mum$,
and $A_{3.3}$ and $A_{4.4}$ are respectively
 the intrinsic band strengths
of the aromatic C--H and C--D stretches
(on a per C--H or C--D bond basis).
Yang et al.\ (2020b) have compiled
all the available observational data
on $\left(I_{3.3}/I_{4.4}\right)_{\rm obs}$
(Peeters et al.\ 2004, Onaka et al.\ 2014, Doney et al.\ 2016)
and obtained a mean value of
$\langle\Iratioobs\rangle\approx52.6$.
With $B_{3.3}/B_{4.4}\approx0.70\pm0.28$
(suitable for $400\simlt T\simlt 900\K$)
and $\ACD/\ACH\approx0.56$
for both mono-deuterated species
(Yang et al.\ 2020b) and multi-deuterated
species (this work), we estimate
$\Dpah\approx2.4\%$.
Compared with the interstellar abundance
of $\Dism\simgt20\pm1\ppm$ (see \S1),
this implies a D enrichment of a factor of
$\simali$1200 in PAHs.

Finally, it is rather puzzling to note that
astronomical observations appear to show
whenever the 4.4$\mum$ aromatic C--D band
was detected, an accompanying aliphatic C--D
band was also present. The latter is often stronger
than the former by a factor of $\simali$2--7
(Peeters et al.\ 2004, Doney et al.\ 2006),
suggesting that a significant amount of D atoms
may be attached to aliphatic C atoms.
While the deuteration of PAHs often requires
UV photons to first rupture the C--H bond
(e.g., see Allamandola et al.\ 1989),
the attachment of aliphatic sidegroups
(whether D-substituted or not) often occurs
in benign environments (e.g., see Yang et al.\ 2017). 
Therefore, it is difficult to understand
how PAHs could be both deuterated and
attached with D-substituted aliphatic sidegroups.
One possibility may be that, in dense molecular
clouds, UV-photolysis of PAHs in D-enriched ices
might lead to deuteration
(i.e., replacing H by D to form aromatic C--D bonds)
{\it and} ``super-deuteration''
(i.e., one H atom and one D atom sharing
a single C atom to form aliphatic C--H and C--D bonds;
e.g., see Bernstein et al.\ 1999, Sandford et al.\ 2000). 
Future experimental and theoretical efforts are
needed for solving this puzzle.
More importantly, future high signal-to-noise
observations by the {\it James Webb Space Telescope}
will provide valuable observational constraints
on the aromatic and aliphatic C--D emission.

\section{Summary}\label{sec:summary}
To facilitate a quantitative understanding
of PAHs as a possible reservoir of interstellar D,
we have employed the hybrid DFT method B3LYP
in conjunction with the 6-311+G$^{\ast\ast}$ basis set
to calculate the vibrational spectra of deuterated
PAHs of different deuterations, sizes and structures
(from pericondened pyrene, perylene, coronene,
and ovalene to linear, catacondensed tetracene).
It is found that the intrinsic band strengths of
the 3.3$\mum$ aromatic C--H stretch ($\Aaro$)
and the 4.4$\mum$ aromatic C--D stretch ($\Acd$)
as well as the band-strength ratios $\Aratio$
do not vary much among PAHs of different deuterations,
sizes and structures. By averaging over all these
(neutral) molecules, we have determined
the mean band strengths to be
$\langle\Aaro\rangle\approx15.3\pm2.1\km\mol^{-1}$
and $\langle\ACD\rangle\approx8.5\pm1.0\km\mol^{-1}$.
Since the mean band-strength ratio of
$\langle \Acd/\Aaro\rangle\approx 0.56\pm0.03$
for {\it multi-deuterated} PAHs is closely similar to
that of {\it mono-deuterated} PAHs,
we conclude that the degree of deuteration
of $\Dpah\approx2.4\%$ {\it previously} estimated
from the observed C--D to C--H intensity ratio
based on the band-strength ratio of
{\it mono-deuterated} PAHs is robust.
Finally, to investigate the charge effect
on the band strengths, we have also calculated
the vibrational spectra of multi-deuterated
pyrene cations and found that, similar to
mono-deuterated PAHs, the C--H and C--D
stretches are substantially suppressed in cations.

\acknowledgments{%
We thank B.T.~Draine, Y.H.~Li
and the anonymous referees
for very helpful suggestions.
XJY is supported in part by NSFC 11873041, 11473023
and the NSFC-CAS Joint Research Funds
in Astronomy (U1731106, U1731107).
AL is supported in part by NASA grants
80NSSC19K0572 and 80NSSC19K0701.
RG is supported in part by NSF-PRISM grant
Mathematics and Life Sciences (0928053).
Computations were performed using the high-performance computer
resources of the University of Missouri Bioinformatics Consortium.
}



\clearpage


\begin{table*}
\footnotesize
\begin{center}
\caption[]{\footnotesize
  Wavelengths($\lambda$)
  and Intensities ($A_\lambda$) 
           of the Nominal ``3.3$\mum$'' C--H Stretch
           and ``4.4$\mum$'' C--D Stretch
           Computed at the B3LYP/6-311+G$^{\ast\ast}$ Level
           for Deuterated Pyrenes of Different Deuterations.
           For Each Deuteration, the Wavelengths and Band Intensities
           Are Obtained by Averaging over
           All the Considered Isomers.
           }
\label{tab:Freq_Int_Pyre_nD_all}
\begin{tabular}{lccccr}
\noalign{\smallskip} \hline \hline \noalign{\smallskip}
Compound	    &	$\lambda_{3.3}$ & $A_{3.3}$
                &	$\lambda_{4.4}$ & $A_{4.4}$  &   $\Aratio$ \\
                &  ($\mu$m)  &  (km$\mol^{-1}$)
                &  ($\mu$m)  & (km$\mol^{-1}$)  &   \\
\noalign{\smallskip} \hline \noalign{\smallskip}
Pyre	    &	3.25 	&	13.97 	&	--   	&	--	    &	--	    \\
Pyre$\_$1D	&	3.25 	&	13.88 	&	4.41 	&	8.24 	&	0.60 	\\
Pyre$\_$2D	&	3.25 	&	14.01 	&	4.40 	&	7.61 	&	0.55 	\\
Pyre$\_$3D	&	3.25 	&	14.00 	&	4.40 	&	7.65 	&	0.55 	\\
Pyre$\_$4D	&	3.25 	&	14.05 	&	4.40 	&	7.62 	&	0.55 	\\
Pyre$\_$5D	&	3.25 	&	14.07 	&	4.40 	&	7.63 	&	0.55 	\\
Pyre$\_$6D	&	3.25 	&	14.09 	&	4.40 	&	7.63 	&	0.55 	\\
Pyre$\_$7D	&	3.26 	&	14.17 	&	4.39 	&	7.62 	&	0.55 	\\
Pyre$\_$8D	&	3.26 	&	14.13 	&	4.39 	&	7.65 	&	0.56 	\\
Pyre$\_$9D	&	3.26 	&	15.11 	&	4.39 	&	7.58 	&	0.53 	\\
Pyre$\_$10D	&	--	    &	--	    &	4.39 	&	7.66 	&	--	    \\
\hline \noalign{\smallskip}
Average 	&	3.26 	&	14.08 	&	4.40 	&	7.63 	&	0.55 	\\
Stdev    	&	0.00 	&	1.18 	&	0.00 	&	0.79 	&	0.09 	\\
\hline
\noalign{\smallskip} \noalign{\smallskip}
\end{tabular}
\end{center}
\end{table*}

\begin{table*}
\footnotesize
\begin{center}
\caption[]{\footnotesize
                Same as Table~\ref{tab:Freq_Int_Pyre_nD_all}
                but for Deuterated Pyrene Cations.
                }
\label{tab:Freq_Int_Pyre_nDPlus_all}
\begin{tabular}{lccccc}
\noalign{\smallskip} \hline \hline \noalign{\smallskip}
Compound	    &	$\lambda_{3.3}$ & $A_{3.3}$
                &	$\lambda_{4.4}$ & $A_{4.4}$  &   $\Aratio$\\
                &  ($\mu$m)  &  (km$\mol^{-1}$)
                &  ($\mu$m)  & (km$\mol^{-1}$)  &   \\
\noalign{\smallskip} \hline \noalign{\smallskip}
Pyre+	    &	3.23 	&	0.24 	&	--  	&	--  	&	--  	\\
Pyre$\_$1D+	&	3.23 	&	0.24 	&	4.37 	&	1.25 	&	5.27 	\\
Pyre$\_$2D+	&	3.24 	&	0.24 	&	4.38 	&	1.45 	&	6.19 	\\
Pyre$\_$3D+	&	3.24 	&	0.24 	&	4.38 	&	1.43 	&	6.06 	\\
Pyre$\_$4D+	&	3.24 	&	0.24 	&	4.38 	&	1.43 	&	6.18 	\\
Pyre$\_$5D+	&	3.24 	&	0.24 	&	4.37 	&	1.42 	&	6.17 	\\
Pyre$\_$6D+	&	3.24 	&	0.24 	&	4.37 	&	1.41 	&	6.31 	\\
Pyre$\_$7D+	&	3.24 	&	0.24 	&	4.37 	&	1.41 	&	6.48 	\\
Pyre$\_$8D+	&	3.24 	&	0.24 	&	4.37 	&	1.40 	&	8.33 	\\
Pyre$\_$9D+	&	3.23 	&	0.21 	&	4.37 	&	1.41 	&	15.26 	\\
Pyre$\_$10D+	&	--	&	--   	&	4.37 	&	1.38 	&	--   	\\
\hline \noalign{\smallskip}
Average 	&	3.24 	&	0.24 	&	4.38 	&	1.42 	&	6.42 	\\
Stdev    	&	0.00 	&	0.04 	&	0.00 	&	0.26 	&	3.45 	\\
\hline
\noalign{\smallskip} \noalign{\smallskip}
\end{tabular}
\end{center}
\end{table*}

\begin{table*}
\footnotesize
\begin{center}
  \caption[]{\footnotesize
    Same as Table~\ref{tab:Freq_Int_Pyre_nD_all}
    but for Deuterated Perylenes.
    }
\label{tab:Freq_Int_Pery_nD_all}
\begin{tabular}{lccccr}
\noalign{\smallskip} \hline \hline \noalign{\smallskip}
Compound	    &	$\lambda_{3.3}$ & $A_{3.3}$
                &	$\lambda_{4.4}$ & $A_{4.4}$  &   $\Aratio$ \\
                &  ($\mu$m)  &  (km$\mol^{-1}$)
                &  ($\mu$m)  & (km$\mol^{-1}$)  &   \\
\noalign{\smallskip} \hline \noalign{\smallskip}
Pery	    &	3.24 	&	13.23 	&	--  	&	--   	&	--  	\\
Pery$\_$1D	&	3.24 	&	13.24 	&	4.39 	&	7.62 	&	0.59 	\\
Pery$\_$2D	&	3.24 	&	13.25 	&	4.39 	&	7.62 	&	0.60 	\\
Pery$\_$3D	&	3.24 	&	13.26 	&	4.39 	&	7.61 	&	0.59 	\\
Pery$\_$4D	&	3.24 	&	12.89 	&	4.39 	&	7.75 	&	0.63 	\\
Pery$\_$5D	&	3.24 	&	13.22 	&	4.39 	&	7.62 	&	0.59 	\\
Pery$\_$6D	&	3.24 	&	13.29 	&	4.39 	&	7.60 	&	0.60 	\\
Pery$\_$7D	&	3.24 	&	13.04 	&	4.39 	&	7.56 	&	0.61 	\\
Pery$\_$8D	&	3.24 	&	13.64 	&	4.39 	&	7.43 	&	0.59 	\\
Pery$\_$9D	&	3.24 	&	13.32 	&	4.39 	&	7.59 	&	0.62 	\\
Pery$\_$10D	&	3.24 	&	13.34 	&	4.39 	&	7.58 	&	0.66 	\\
Pery$\_$11D	&	3.24 	&	13.35 	&	4.39 	&	7.58 	&	0.71 	\\
Pery$\_$12D	&	--  	&	--   	&	4.39 	&	7.58 	&	--   	\\
\hline \noalign{\smallskip}
Average   	&	3.24 	&	13.14 	&	4.39 	&	7.62 	&	0.61 	\\
Stdev	    &	0.00 	&	2.09 	&	0.01 	&	1.56 	&	0.22 	\\
\hline
\noalign{\smallskip} \noalign{\smallskip}
\end{tabular}
\end{center}
\end{table*}

\begin{table*}
\footnotesize
\begin{center}
  \caption[]{\footnotesize
    Same as Table~\ref{tab:Freq_Int_Pyre_nD_all}
    but for Deuterated Coronenes.
           }
\label{tab:Freq_Int_Coro_nD_all}
\begin{tabular}{lccccr}
\noalign{\smallskip} \hline \hline \noalign{\smallskip}
Compound	    &	$\lambda_{3.3}$ & $A_{3.3}$
                &	$\lambda_{4.4}$ & $A_{4.4}$  &   $\Aratio$ \\
                &  ($\mu$m)  &  (km$\mol^{-1}$)
                &  ($\mu$m)  & (km$\mol^{-1}$)  &   \\
\noalign{\smallskip} \hline \noalign{\smallskip}
Coro	    &	3.26 	&	16.17 	&	--	    &	--  	&	-- 	    \\
Coro$\_$1D	&	3.26 	&	16.19 	&	4.41 	&	8.95 	&	0.55 	\\
Coro$\_$2D	&	3.26 	&	16.21 	&	4.41 	&	8.97 	&	0.55 	\\
Coro$\_$3D	&	3.26 	&	16.24 	&	4.41 	&	8.97 	&	0.55 	\\
Coro$\_$4D	&	3.26 	&	16.26 	&	4.41 	&	8.98 	&	0.55 	\\
Coro$\_$5D	&	3.26 	&	16.28 	&	4.41 	&	8.99 	&	0.55 	\\
Coro$\_$6D	&	3.26 	&	16.30 	&	4.41 	&	9.00 	&	0.55 	\\
Coro$\_$7D	&	3.26 	&	16.32 	&	4.41 	&	9.02 	&	0.55 	\\
Coro$\_$8D	&	3.26 	&	16.34 	&	4.41 	&	9.03 	&	0.55 	\\
Coro$\_$9D	&	3.26 	&	16.37 	&	4.41 	&	9.04 	&	0.55 	\\
Coro$\_$10D	&	3.26 	&	16.38 	&	4.41 	&	9.05 	&	0.55 	\\
Coro$\_$11D	&	3.26 	&	16.41 	&	4.41 	&	9.06 	&	0.55 	\\
Coro$\_$12D	&	--	    &	--   	&	4.41 	&	9.07 	&	--  	\\
\hline \noalign{\smallskip}
Average  	&	3.26 	&	16.30 	&	4.41 	&	9.00 	&	0.55 	\\
Stdev  	    &	0.00 	&	0.06 	&	0.00 	&	0.03 	&	0.00 	\\
\hline
\noalign{\smallskip} \noalign{\smallskip}
\end{tabular}
\end{center}
\end{table*}

\begin{table*}
\footnotesize
\begin{center}
  \caption[]{\footnotesize
    Same as Table~\ref{tab:Freq_Int_Pyre_nD_all}
    but for Deuterated Ovalenes.
           }
\label{tab:Freq_Int_Oval_nD_all}
\begin{tabular}{lccccr}
\noalign{\smallskip} \hline \hline \noalign{\smallskip}
Compound	    &	$\lambda_{3.3}$ & $A_{3.3}$
                &	$\lambda_{4.4}$ & $A_{4.4}$  &   $\Aratio$ \\
                &  ($\mu$m)  &  (km$\mol^{-1}$)
                &  ($\mu$m)  & (km$\mol^{-1}$)  &   \\
\noalign{\smallskip} \hline \noalign{\smallskip}
Oval	    &	3.26 	&	17.68 	&	--   	&	--   	&	--   	\\
Oval$\_1$D	&	3.26 	&	17.65 	&	4.41 	&	10.16 	&	0.58 	\\
Oval$\_2$D	&	3.26 	&	17.73 	&	4.41 	&	9.90 	&	0.56 	\\
Oval$\_3$D	&	3.26 	&	17.59 	&	4.41 	&	8.94 	&	0.51 	\\
Oval$\_4$D	&	3.26 	&	17.68 	&	4.41 	&	10.12 	&	0.57 	\\
Oval$\_5$D	&	3.26 	&	17.76 	&	4.41 	&	10.00 	&	0.56 	\\
Oval$\_6$D	&	3.26 	&	17.94 	&	4.41 	&	9.30 	&	0.52 	\\
Oval$\_7$D	&	3.26 	&	17.78 	&	4.41 	&	10.03 	&	0.56 	\\
Oval$\_8$D	&	3.26 	&	17.82 	&	4.41 	&	10.01 	&	0.56 	\\
Oval$\_9$D	&	3.26 	&	17.87 	&	4.41 	&	9.73 	&	0.55 	\\
Oval$\_10$D	&	3.26 	&	17.97 	&	4.41 	&	9.98 	&	0.56 	\\
Oval$\_11$D	&	3.26 	&	17.84 	&	4.41 	&	10.04 	&	0.57 	\\
Oval$\_12$D	&	3.26 	&	17.85 	&	4.41 	&	10.05 	&	0.57 	\\
Oval$\_13$D	&	3.26 	&	18.63 	&	4.41 	&	10.04 	&	0.55 	\\
Oval$\_14$D	&	--	    &	--	    &	4.41 	&	10.07 	&	--	\\
\hline \noalign{\smallskip}
Average	    &	3.26 	&	17.83 	&	4.41 	&	9.67 	&	0.55 	\\
Stdev	    &	0.00 	&	0.82 	&	0.00 	&	1.00 	&	0.07 	\\
\hline
\noalign{\smallskip} \noalign{\smallskip}
\end{tabular}
\end{center}
\end{table*}

\begin{table*}
\footnotesize
\begin{center}
  \caption[]{\footnotesize
    Same as Table~\ref{tab:Freq_Int_Pyre_nD_all}
    but for Deuterated Tetracenes.
           }
\label{tab:Freq_Int_Tetr_nD_all}
\begin{tabular}{lccccr}
\noalign{\smallskip} \hline \hline \noalign{\smallskip}
Compound	    &	$\lambda_{3.3}$ & $A_{3.3}$
                &	$\lambda_{4.4}$ & $A_{4.4}$  &   $\Aratio$ \\
                &  ($\mu$m)  &  (km$\mol^{-1}$)
                &  ($\mu$m)  & (km$\mol^{-1}$)  &   \\
\noalign{\smallskip} \hline \noalign{\smallskip}
Tetr	    &	3.26 	&	13.59 	&	--  	&	--  	&	--  	\\
Tetr$\_$1D	&	3.26 	&	13.61 	&	4.41 	&	7.50 	&	0.55 	\\
Tetr$\_$2D	&	3.26 	&	13.63 	&	4.41 	&	7.51 	&	0.55 	\\
Tetr$\_$3D	&	3.26 	&	13.65 	&	4.41 	&	7.52 	&	0.55 	\\
Tetr$\_$4D	&	3.26 	&	13.56 	&	4.41 	&	7.16 	&	0.53 	\\
Tetr$\_$5D	&	3.26 	&	13.50 	&	4.41 	&	7.61 	&	0.57 	\\
Tetr$\_$6D	&	3.26 	&	13.70 	&	4.41 	&	7.55 	&	0.55 	\\
Tetr$\_$7D	&	3.26 	&	13.79 	&	4.41 	&	7.28 	&	0.53 	\\
Tetr$\_$8D	&	3.26 	&	12.60 	&	4.41 	&	7.70 	&	0.62 	\\
Tetr$\_$9D	&	3.26 	&	13.75 	&	4.41 	&	7.59 	&	0.56 	\\
Tetr$\_$10D	&	3.26 	&	13.77 	&	4.41 	&	7.60 	&	0.57 	\\
Tetr$\_$11D	&	3.26 	&	13.78 	&	4.41 	&	7.61 	&	0.58 	\\
Tetr$\_$12D	&	--  	&	--  	&	4.41 	&	7.62 	&	--  	\\
\hline \noalign{\smallskip}
Average	   &	3.26 	&	13.64 	&	4.41 	&	7.40 	&	0.55 	\\
Stdev  	   &	0.00 	&	1.01 	&	0.01 	&	0.54 	&	0.07 	\\
\hline
\noalign{\smallskip} \noalign{\smallskip}
\end{tabular}
\end{center}
\end{table*}


\begin{thebibliography}{30}
\expandafter\ifx\csname natexlab\endcsname\relax\def\natexlab#1{#1}\fi
%

\bibitem[]{}Allamandola, L.J., Tielens, A.G.G.M., \& Barker, J.R.\
            1985, ApJ, 290, L25
\bibitem{}Allamandola, L.~J., Sandford, S.~A., \& Wopenka, B.\
                 1987, Science, 237, 56
\bibitem{}Allamandola, L.J., Tielens, A.G.G.M., \& Barker, J.R.\
            1989, ApJS, 71, 733
\bibitem{}Allamandola, L.J., Hudgins, D.M., \& Sandford, S.A.\
            1999, ApJ, 511, 115
\bibitem{}Allen, M. M., Jenkins, E. B., \& Snow, T. P.\
            1992, ApJS, 83, 261
\bibitem{}Bauschlicher, C. W.,  Langhoff, S. R.,
          Sandford, S. A., \& Hudgins, D. M.\ 1997,
          J. Phys. Chem. A, 101, 2414
\bibitem{}Bauschlicher, C.~W., Peeters, E.,
          \& Allamandola, L.~J.\ 2008, ApJ, 678, 316
\bibitem{}Bauschlicher, C.~W., Peeters, E.,
  \& Allamandola, L.~J.\ 2009, ApJ, 697, 311
\bibitem{}Bauschlicher, C.~W., Ricca, A.,
               Boersma, C., Allamandola, L.~J.\
               2018, ApJS, 234, 32
\bibitem[]{}Bernstein, L.~S., Shroll, R.~M.,
                  Lynch, D.~K., \& Clark, F.~O.\
                  2017, ApJ, 836, 229
\bibitem[]{}Bernstein, M.P., Sandford, S.A., \& Allamandola, L.J.\
            1996, ApJ, 472, L127
\bibitem{}Bernstein, M. P., Sandford, S. A., Allamandola, L. J., et al.\ 1999,
          Science, 283, 1135
\bibitem{}Boesgaard, A. M., \& Steigman, G.\ 1985,
          ARA\&A, 23, 319
\bibitem{}Borowski, P.\ 2012,
            J. Phys. Chem. A, 116, 3866
\bibitem{}Buragohain, M., Pathak, A., Sarre, P., Onaka, T., \& Sakon, I.\ 2015,
          MNRAS, 454, 193
\bibitem{}Buragohain, M., Pathak, A., Sarre, P.,
                  Onaka, T., \& Sakon, I.\
                  2016, Planet. Space Sci., 133, 97
\bibitem{}Buragohain, M., Pathak, A., Sakon, I.,
               \& Onaka, T.\ 2020, ApJ, 892, 11
\bibitem{}Cami, J.\ 2011, EAS Publ. Ser. 46,
  PAHs and the Universe: A Symposium to
  Celebrate the 25th Anniversary of
  the PAH Hypothesis,
  ed. C. Joblin \& A.G.G.M. Tielens
  (Les Ulis: EDP Sciences), 117
\bibitem{}Coc, A., Vangioni-Flam, E., Descouvemont, P.,
          Adahchour, A., \& Angulo, C.\ 2004,
          ApJ, 600, 544
\bibitem{}Cooke, R., Pettini, M., \& Steidel, Charles C.\ 2018,
          ApJ, 855, 102
\bibitem{}Cruz-Diaz, G.A., Ricca, A., \& Mattioda, A.L.\
                2020, ACS Earth Space Chem., 4, 1730
\bibitem{}Doney, K. D., Candian, A., Mori, T., Onaka, T.,
          \& Tielens, A. G. G. M.\ 2016,
          A\&A, 586, 65
\bibitem{}Draine, B.~T.\ 2004,
                in Origin and Evolution of the Elements,
                ed. A. McWilliam \& M. Rauch
                (Cambridge: Cambridge Univ. Press), 317
\bibitem{}Draine, B.~T.\ 2006,
  in ASP Conf. Ser. 348,
  Astrophysics in the Far Ultraviolet:
  Five Years of Discovery with FUSE,
  ed. G. Sonneborn, H. Moos, \& B.-G. Andersson
  (San Francisco, CA: ASP), 58
\bibitem[]{}Draine, B.T., \& Li, A.\ 2007, ApJ, 657, 810
\bibitem[]{}Draine, B.T., Li, A., Hensley, B.S., Hunt, L.K.,
                  Sandstrom, K., \& Smith, J.-D.T.\
                  2021, ApJ, in press (arXiv:2011.07046)
\bibitem{}Frisch, M. J., Trucks, G. W., Schlegel, H. B.,
            et al.\ 2016, Gaussian 16, Revision C. 01,
            Gaussian, Inc., Wallingford CT
\bibitem[]{}Hudgins, D. M., \& Allamandola, L. J.\ 
            2005, in IAU Symp.\,231, 
            Astrochemistry: Recent Successes 
            and Current Challenges,
            ed. D. C. Lis, G. A. Blake, \& E. Herbst 
            (Cambridge: Cambridge Univ. Press), 443
\bibitem{}Hudgins, D. M., Sandford, S. A.,
                Allamandola, L. J.\ 1994,
                J. Phys. Chem., 98, 4243
\bibitem{}Hudgins, D.~M., Bauschlicher, C.~W., Jr.,
                \& Sandford, S.~A.\ 2004,
                ApJ, 614, 770
                
\bibitem{}Hudgins, D. M., Bauschlicher, C. W. Jr.,
                 \& Allamandola, L. J.\ 2005, ApJ, 632, 316

\bibitem{}Jenkins, E. B., Tripp, T. M., W{\'o}zniak, P., Sofia, U. J., \& Sonneborn, G.\ 1999,
          ApJ, 520,182

\bibitem{}Kerridge, J. F., Chang, S. \& Shipp, R.\ 1987,
          GeCoA, 51, 2527

\bibitem{}Kwok, S., \& Zhang, Y.\ 2011, Nature, 479, 80
\bibitem{}Kwok, S., \& Zhang, Y.\ 2013, ApJ, 771, 5
  
\bibitem{}Li, A.\ 2020, Nature Astronomy, 4, 339

\bibitem[]{}Li, A., \& Draine, B.~T.\ 2001, ApJ, 554, 778
  
\bibitem{}Loinard, L., Castets, A., Ceccarelli, C., et al.\ 2000,
          A\&A, 359, 1169L

\bibitem{}Mattioda, A.~L., Rutter, L., Parkhill, J.,
                 Head-Gordon, M., Lee, T.J.,
                 \& Allamandola, L.J.\ 2008, ApJ, 680, 1243
              
\bibitem{}Mattioda, A.~L., Hudgins, D.~M.,
                Boersma, C., et al.\ 2020, ApJS, 251, 22.
  
\bibitem{}Mazzitelli, I., \& Moretti, M.\ 1980,
               ApJ, 235, 995
  
\bibitem{}McGuire, B.~A., Loomis, R.~A.,
                 Burkhardt, A.~M., et al.\
                 2021, Science, 371, 1265

\bibitem{}Onaka, T., Mori, T.~I., Sakon, I.,
                et al.\ 2014,
                ApJ, 780, 114

\bibitem{}Parise, B., Ceccarelli, C., Tielens, A. G. G. M., et al.\ 2002,
          A\&A, 393, L49


\bibitem{}Peeters, E., Allamandola, L.~J.,
                 Bauschlicher, C.~W., Jr., et al.\ 2004,
                 ApJ, 604, 252


\bibitem{}Prodanovi{\'c}, T., Steigman, G., \& Fields, B. D.\ 2010,
          MNRAS, 406, 1108

\bibitem{}Rosenberg, M.~J.~F., Bern{\'e}, O., Boersma, C.,
               Allamandola, L. J., \& Tielens, A. G. G. M.\ 
               2011, A\&A, 532, A128
             
\bibitem{}Roueff, E., Tin, S., Coudert, L. H., et al.\ 2000,
               A\&A, 354, L63
  
\bibitem{}Sadjadi, S., Zhang, Y., \& Kwok, S.\
                2017, ApJ, 845, 123. 

          
\bibitem{}Sakata, A., Wada, S., Onaka, T.,
                 \& Tokunaga, A.T.\ 1990, ApJ, 353, 543
          
\bibitem{}S{\'a}nchez, Ariel G., Baugh, C. M., Percival, W. J., et al.\ 2006,
          MNRAS, 366, 189

\bibitem[]{}Sandford, S.A.\ 1991, ApJ, 376, 599

\bibitem{}Sandford, S.A., Bernstein, M.P.,
                  \& Dworkin, J.P.\ 2001,
                  Meteoritics Planet. Sci., 36, 1117

\bibitem{}Sandford, S. A., Allamandola, L. J.,
               Tielens, A. G. G. M., et al.\ 1991,
          ApJ, 371, 607

\bibitem{}Sandford, S.A., Bernstein, M.~P.,
                 \& Materese, C.K.\
                2013, ApJS, 205, 8

\bibitem{}Spergel, D. N., Verde, L., Peiris, H. V., et al.\ 2003,
          ApJS, 148, 175

\bibitem{}van der Tak, F.F.S., Schilke, P., M{\"{u}}uller, H.S.P., et al.\ 2002,
          A\&A, 388, L53

\bibitem[]{}Thrower, J. D., J{\o}rgensen, B., Friis, E. E., et al.\
            2012, ApJ, 752, 3
          
\bibitem{}Vastel, C., Phillips, T. G., Ceccarelli, C., \& Pearson, J.\ 2003,
          ApJL, 593, L97

\bibitem{}Wexler, A.S.\ 1967, Appl. Spectro. Rev., 1, 29

\bibitem{}Yang, X.~J., Glaser, R., Li, A.,
                  \& Zhong, J.~X.\
                  2013, ApJ, 776, 110
                  
\bibitem{}Yang, X.~J., Li, A., \& Glaser, R.\
                2016, ApJ, 825, 22

\bibitem{}Yang, X.~J., Glaser, R., Li, A.,
                  \& Zhong, J.~X.\
                 2017, New Astron. Rev., 77, 1

\bibitem{}Yang, X.~J., Li, A., \& Glaser, R.\
            2020a, ApJS, 247, 1

\bibitem{}Yang, X.~J., Li, A., \& Glaser, R.\
            2020b, ApJS, 251, 12

\bibitem{}Zavarygin, E. O., Webb, J. K.,
              Riemer-S{\o}rensen, S., \& Dumont, V.\ 2018,
              MNRAS, 477, 5536

\bibitem{}Zhang, Y., \& Kwok, S.\ 2015, 
               ApJ, 798, 37
%
\end{thebibliography}
\end{document}